\documentclass[lettersize,journal]{IEEEtran}
\def\BibTeX{{\rm B\kern-.05em{\sc i\kern-.025em b}\kern-.08em
		T\kern-.1667em\lower.7ex\hbox{E}\kern-.125emX}}
\usepackage{cite}
\usepackage{algpseudocode}
\usepackage{graphicx}
\usepackage{amsmath}
\usepackage{amsfonts}
\usepackage{epstopdf}
\usepackage{xcolor}
\usepackage{enumerate}
\usepackage{algorithm}
\usepackage{multirow}
\usepackage{transparent}
\usepackage{subfigure}
\usepackage{float}

\newtheorem{theorem}{Theorem}
\newtheorem{remark}{Remark}
\newtheorem{corollary}{Corollary}
\newtheorem{lemma}{Lemma}

\newcommand{\T}{\textnormal{T}}
\newcommand{\R}{\textnormal{R}}
\newcommand{\s}{\textnormal{s}}

\let\ss= \scriptscriptstyle
\begin{document}
	\title{Membrane Fusion-Based Transmitter Design for Static and Diffusive Mobile Molecular Communication Systems}
	\author{\thanks{This work was presented in part at 2021 IEEE International Conference on Communication (ICC) \cite{1907.04239}.}Xinyu Huang\thanks{X. Huang, N. Yang are with the School of Engineering, Australian National University, Canberra, ACT 2600, Australia (e-mail: \{xinyu.huang1, nan.yang\}@anu.edu.au).}, \textit{Student Member, IEEE}, Yuting Fang\thanks{Y. Fang is with the Department of Electrical and Electronic Engineering, University of Melbourne, Parkville, VIC 2010, Australia (e-mail: yuting.fang@unimelb.edu.au).}, Adam Noel, \textit{Member, IEEE}\thanks{A. Noel is with the School of Engineering, University of Warwick, Coventry, CV 7AL, UK (e-mail: adam.noel@warwick.ac.uk).}, \\and Nan Yang, \textit{Senior Member, IEEE}\vspace{-8mm}}
	
	\maketitle
	
	\begin{abstract}
		This paper proposes a novel imperfect transmitter (TX) model, namely the membrane fusion (MF)-based TX, that adopts MF between a vesicle and the TX membrane to release molecules encapsulated within the vesicle. For the MF-based TX, the molecule release probability and the fraction of molecules released from the TX membrane are derived. Incorporating molecular degradation and a fully-absorbing receiver (RX), the channel impulse response (CIR) is derived for two scenarios: 1) Both TX and RX are static, and 2) both TX and RX are diffusion-based mobile. Moreover, a sequence of bits transmitted from the TX to the RX is considered. The average bit error rate (BER) is obtained for both scenarios, wherein the probability mass function (PMF) of the number of molecules absorbed in the mobile scenario is derived. Furthermore, a simulation framework is proposed for the MF-based TX, based on which the derived analytical expressions are validated. Simulation results show that a low MF probability or low vesicle mobility slows the release of molecules and reduces the molecule hitting probability at the RX. Simulation results also indicate the difference between the MF-based TX and an ideal point TX in terms of the inter-symbol interference (ISI).
	\end{abstract}
	
	
	\section{Introduction}
	With the rapid development of nano-technology, communication between nano-machines has become a widely-investigated problem. Inspired by nature, one promising solution to this problem is to use chemical signals as information carriers, which is referred to as molecular communication (MC) \cite{nakano2013molecular}. In MC, a transmitter (TX) releases molecules into a fluid medium, where the molecules propagate until they arrive at a receiver (RX). An end-to-end MC system model consists of a TX, a propagation environment, and a RX as in a conventional wireless system. Molecule propagation environments and reception mechanisms at the RX have been widely investigated in previous studies \cite{jamali2019channel}. However, few studies have investigated the impact on MC system performance by the signaling pathways \emph{inside} the TX and the interaction of molecular signals with the TX surface.
	
	Most existing studies have assumed the TX to be an ideal point source that can release molecules instantaneously \cite{jamali2019channel}. Compared to realistic scenarios, this ideal TX ignores TX properties including geometry, signaling pathways inside the TX, and chemical reactions during the release process. Recently, some studies considered these TX properties, e.g., \cite{yilmaz2017chemical,arjmandi2016ion,schafer2019spherical}.\cite{yilmaz2017chemical} proposed a spherical TX whose boundary reflects the emitted molecules and investigated the directivity gain achieved by the reflecting TX, where the directivity gain measures the degree to which the emitted molecules are concentrated in a single direction. \cite{arjmandi2016ion} proposed an ion channel-based TX, where molecule release is controlled by the opening and closing of ion channels. \cite{schafer2019spherical} considered a spherical TX with a semi-permeable boundary whose permeability is used to control molecule release. Although these studies stand on their own merits, none of them considered a chemical reaction-driven process for molecule release.
	
	In nature, exocytosis is a form of active transport in which a cell transports molecules out of the cell by secreting them through an energy-dependent process \cite{jahn1999membrane}. Exocytosis is common for cells because many chemical substances are large molecules, e.g., neurotransmitters and proteins, and cannot pass through the cell membrane by passive means \cite{biomor}. In exocytosis, vesicles\footnote{A vesicle is a small, round or oval-shaped container for the storage of molecules and as compartments for particular chemical reactions \cite{bonifacino2004mechanisms}.} are carried to the cell membrane to secrete their contents into the extracellular environment. This secretion is performed by the membrane fusion (MF) process that fuses the vesicle with the cell membrane. When the vesicle moves close to the cell membrane, the v-SNARE proteins on the vesicle membrane bind to the t-SNARE proteins on the cell membrane, which generates the \textit{trans}-SNARE complexes that catalyze MF \cite{bonifacino2004mechanisms}. Unlike the ideal point TX, exocytosis is an existing biological process and constrains the perfect release of molecules. Moreover, a TX model that is designed based on the exocytosis process has many applications in both natural and synthetic MC systems. The applications of the TX model in the natural MC system include facilitating the communication between cells by releasing signaling molecules, modeling the transportation process of hormones insulin and glucagon from the pancreas to control the glucose concentration in the blood, and investigating the process of removing toxins or waste products from the cell's interior to maintain homeostasis \cite{wu2014exocytosis}. As the TX model captures some basic functions of the cell, it can be designed by modifying biological cells \cite{nakano2012molecular}, which improves the degree of biocompatibility for the TX model in the human body for targeted drug delivery. Specifically, drug molecules are stored within the extracellular vesicles that are released from the MF-based TX by the exocytosis process \cite{de2019drug}. Therefore, a TX that uses MF to release molecules merits investigation.
	
	Recently, mobile TXs and RXs in MC have drawn considerable attention since they are required in many applications, e.g., targeted drug delivery, water quality control \cite{akyildiz2011nanonetworks}, and human body monitoring \cite{huang2020parameter}. As the distance between the TX and the RX in such cases can be continuously changing while transmitting information, the channel is stochastic and it is challenging to perform statistical analysis of the received signal at the RX. Unknown statistical properties, e.g., mean and probability mass function (PMF), of the received signal bring difficulties in channel performance analysis. Motivated by this, some studies have investigated mobile MC and established a stochastic framework for channel modeling, e.g., \cite{ahmadzadeh2018stochastic,cao2019diffusive,huang2019statistical}. However, these studies considered a mobile ideal point TX that releases molecules instantaneously while moving. \cite{ahmadzadeh2018stochastic} and \cite{cao2019diffusive} considered a mobile transparent RX that does not interact with molecules and a mobile non-transparent RX, respectively, where statistical properties of the time-varying channel were derived. For bit transmission, \cite{ahmadzadeh2018stochastic} assumed that the distance between the TX and the RX is known when bits are transmitted, and \cite{cao2019diffusive} ignored the inter-symbol interference (ISI) when performing the error probability analysis. \cite{huang2019statistical} derived the distribution of the number of molecules observed at a mobile transparent RX. To the best of the authors' knowledge, no studies have considered stochastic channel modeling between a \emph{mobile non-ideal TX} and a \emph{mobile non-transparent RX}. Furthermore, the statistical analysis of the received signal when a mobile TX emits molecules \emph{non-instantaneously} has not been investigated.
	
	In this paper, we propose a novel TX model in a three-dimensional (3D) environment, namely the MF-based TX, which uses fusion between a vesicle generated within the TX and the TX membrane to release molecules encapsulated within the vesicle. By considering a fully-absorbing RX that absorbs molecules once they hit the RX surface, we investigate the end-to-end channel impulse response (CIR) between the MF-based TX and the RX in two scenarios: 1) A static TX and a static RX, and 2) a diffusion-based mobile TX and a diffusion-based mobile RX, where the CIR is the molecule hitting probability at the RX when an impulse of vesicles are released from the TX's center \cite{jamali2019channel}. Moreover, we consider a sequence of bits transmitted from the TX to the RX and investigate the error probability at the RX for both scenarios. For the mobile scenario, the distance between the TX and RX constantly changes during the non-instantaneous molecule release process such that the statistical analysis of the received signal at the mobile RX is more challenging than that in the instantaneous molecule release process.
	To tackle this challenge, we propose a novel method that divides the molecule release process into multiple intervals to calculate the PMF of the number of absorbed molecules. In summary, our major contributions are as follows:
	\begin{itemize}
		\item[1)] We derive the time-varying molecule release probability and the fraction of molecules released from the TX by a given time.
		\item[2)] For the static scenario, we derive the end-to-end i) molecule hitting probability, ii) the fraction of molecules absorbed, and iii) the asymptotic fraction of molecules absorbed at the RX due to the MF-based TX as time approaches infinity. For the mobile scenario, we derive the expected end-to-end molecule hitting probability at the mobile RX.
		\item[3)] By considering a sequence of bits transmitted from the TX and a threshold detector at the RX, we calculate the average bit error rate (BER) for the static and mobile scenarios. To derive the average BER in the mobile scenario, we also derive the PMF for the number of molecules absorbed at the mobile RX. 
		\item[4)] We propose a simulation framework for the MF-based TX model to simulate the diffusion and fusion of vesicles within the TX. In this simulation framework, the release point on the TX membrane and the release time of each molecule are determined.
	\end{itemize}
	Aided by the proposed simulation framework, we demonstrate the accuracy of our analytical derivations. Our numerical results show that a low MF probability or low vesicle mobility slows the release of molecules from the TX, increases the time to reach the peak molecule release probability, and reduces the end-to-end molecule hitting probability at the RX. Moreover, numerical results show the difference between the MF-based TX and an ideal point TX from the perspective of ISI. For the ideal point TX, ISI is only caused by the diffusion of molecules in the propagation environment. For the MF-based TX, ISI can also be introduced by signaling pathways inside the TX.
	
	This paper extends our preliminary work in \cite{1907.04239} as follows. First, we add transceiver mobility while \cite{1907.04239} only considered the static scenario. Second, we add the analysis of the end-to-end fraction of molecules absorbed and the asymptotic fraction of molecules absorbed at the RX for the static scenario. Third, we add channel performance analysis by considering the transmission of a bit sequence in both scenarios.
	
	The rest of this paper is organized as follows. In Section \ref{SM}, we introduce the system model. In Section \ref{dCIRs}, we derive the release probability and fraction of molecules released from the TX. We also derive the end-to-end hitting probability, fraction of molecules absorbed, and asymptotic fraction of molecules absorbed at the RX for the static scenario. In Section \ref{DCIRm}, we derive the expected end-to-end hitting probability for the mobile scenario. In Section \ref{sber}, we derive the average BER for both scenarios. For the mobile scenario, we also derive and verify the PMF of absorbed molecules. In Section \ref{SF}, we propose a simulation framework for the MF-based TX. In Section \ref{NR}, we discuss the numerical results, and conclusions are presented in Section \ref{con}.
	
	\section{System Model}\label{SM}
	\begin{figure*}[!t]
		\begin{center}
			\includegraphics[height=3.4in,width=1.9\columnwidth]{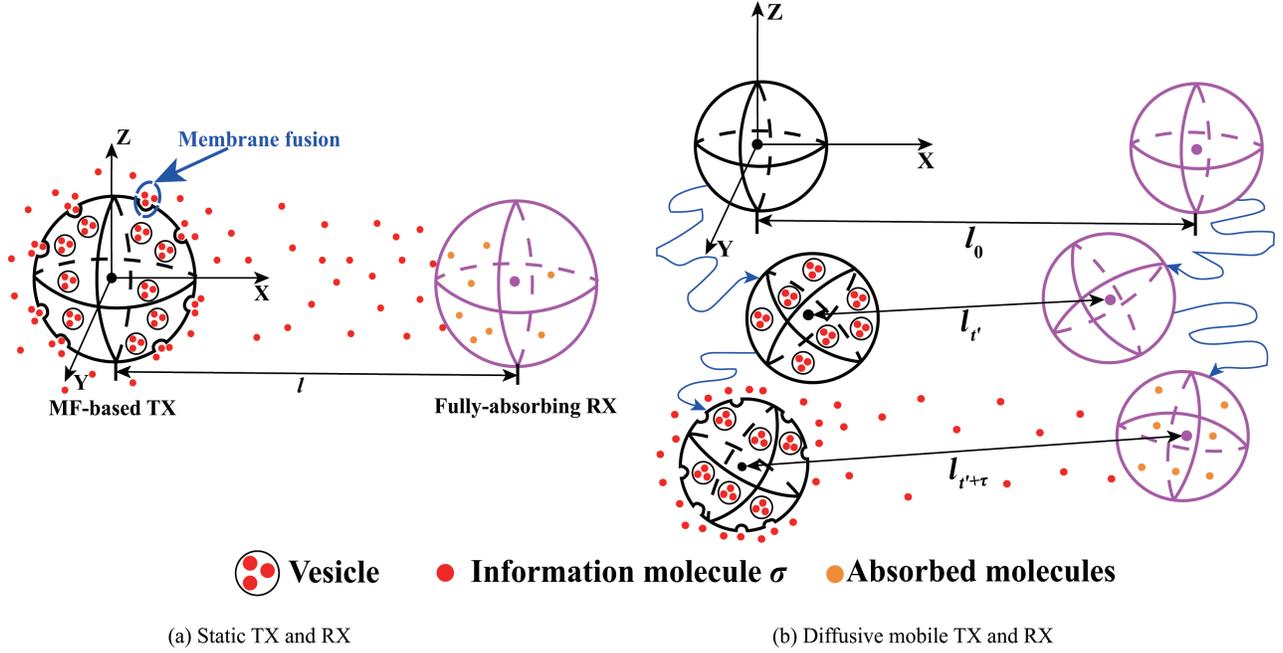}
			\caption{Illustration of the system model where one MF-based TX communicates with one fully-absorbing RX in a 3D environment. Fig. 1(a) shows static TX and RX such that the TX-RX distance is fixed and Fig. 1(b) shows diffusive mobile TX and RX such that the TX-RX distance changes over time.}\label{sys}\vspace{-0.5em}
		\end{center}
		\vspace{-4mm}
	\end{figure*}
	In this paper, we consider an unbounded 3D environment where an MF-based TX communicates with a fully-absorbing RX, as depicted in Fig. \ref{sys}. We investigate two scenarios: 1) A static TX and a static RX, and 2) a diffusive mobile TX and a diffusive mobile RX, as depicted in Figs. 1(a) and 1(b), respectively. Both TX and RX are spheres with radius $r_{\ss\T}$ and $r_{\ss\R}$, respectively. We assume that the propagation environment outside the spherical TX and RX is the fluid medium with uniform temperature and viscosity. Once information molecules $\sigma$ are released from the TX, they diffuse randomly with a constant diffusion coefficient $D_{\sigma}$. Moreover, we consider unimolecular degradation in the propagation environment, where type $\sigma$ molecules can degrade into type $\hat{\sigma}$ molecules that cannot be identified by the RX, i.e., $\sigma\stackrel{k_\mathrm{d}}{\longrightarrow}\hat{\sigma}$ \cite[Ch. 9]{chang2005physical}, and $k_\mathrm{d}\; \left[\mathrm{s}^{-1}\right]$ is the degradation rate. In addition, we model the RX as a fully-absorbing RX by which molecules $\sigma$ are absorbed once they hit the RX surface.
	\subsection{MF-based TX Model}\label{MF}
	We assume that the spherical TX releases molecules from its outer membrane after fusion between the membrane and vesicles. Each vesicle stores and transports $\eta$ type $\sigma$ molecules. We consider that the TX is filled with the fluid medium that has uniform temperature and viscosity. Generally, vesicles can diffuse in the fluid medium based on the experiment in \cite{kyoung2008vesicle} or can be actively transported along microtubules by motor proteins \cite{sheetz1987movements}. In this paper, we focus on the diffusion of vesicles. We assume that a vesicle cluster \cite{milovanovic2017synaptic} exists within the TX. Vesicle clustering is a biological phenomenon observed in the presynaptic neuron, where vesicles can form clusters to sustain a high release probability of neurotransmitters. Vesicles can be released from the cluster via dephosphorylation \cite{henkel1996synaptic}. Since we assume that the size of a vesicle \cite{cosson2002resident} is relatively small compared to the size of the TX, we mathematically model the vesicle release process as an instantaneous release from a point within the TX. Once vesicles are released, they diffuse randomly with a constant diffusion coefficient $D_\mathrm{v}$. According to \cite{bonifacino2004mechanisms}, the natural fusion of a vesicle and the cell membrane can be considered as two steps: 1) v-SNARE proteins ($\mathrm{S}_\mathrm{v}$) on the vesicle membrane bind to t-SNARE proteins ($\mathrm{S}_\mathrm{t}$) on the cell membrane to generate \textit{trans}-SNARE complexes ($\mathrm{S}_\mathrm{c}$), and 2) $\mathrm{S}_\mathrm{c}$ catalyzes the fusion of vesicular and cell membranes. For tractability, we apply the following main assumptions to the TX model:
	\begin{itemize}
		\item[A1)] Vesicles are released from the TX's center. This assumption ensures that molecules are uniformly released from the TX membrane. Considering vesicles released from any point within the TX is an interesting topic for future work. 
		\item[A2)] The interaction between $\mathrm{S}_\mathrm{v}$ and $\mathrm{S}_\mathrm{t}$ is modeled as the irreversible reaction $\mathrm{S}_\mathrm{v}+\mathrm{S}_\mathrm{t}\rightarrow\mathrm{S}_\mathrm{c}$, which indicates that we only focus on the forward reaction between $\mathrm{S}_\mathrm{v}$ and $\mathrm{S}_\mathrm{t}$. The backward reaction for the disassembly of $\mathrm{S}_\mathrm{c}$ occurs after MF, wherein the free $\mathrm{S}_\mathrm{v}$ and $\mathrm{S}_\mathrm{t}$ are used for the subsequent rounds of transport \cite{hong2005snares}. This is beyond the scope of our system model. After $\mathrm{S}_\mathrm{c}$ is generated, it catalyzes MF, which is $\mathrm{S}_\mathrm{c}\rightarrow\mathrm{MF}$ \cite{hua2001three}. We simplify this reaction network by removing the intermediate product $\mathrm{S}_\mathrm{c}$ and combining these two reactions \cite{madelaine2017simplification}, which is given by
		\begin{align}\label{IR}
			\mathrm{S}_\mathrm{v}+\mathrm{S}_\mathrm{t}\stackrel{k_\mathrm{f}}{\longrightarrow} \mathrm{MF},
		\end{align}
		where $k_\mathrm{f}$ is the forward reaction rate in $\mu\mathrm{m}/\mathrm{s}$.
		
		\item[A3)] The membrane is fully covered by an infinite number of $\mathrm{S}_\mathrm{t}$ and the occupancy of $\mathrm{S}_\mathrm{t}$ by $\mathrm{S}_\mathrm{v}$ is ignored. Assuming perfect receptor coverage and ignoring occupancy are for the sake of tractability and 
		have been widely adopted in previous studies, e.g., \cite{arjmandi2019diffusive,ahmadzadeh2016comprehensive}.
		\item[A4)] Once molecules are released, the spherical TX is assumed to not hinder the random diffusion of molecules in the propagation environment, i.e, the TX is transparent to the diffusion of released molecules. \cite{yilmaz2017chemical} analyzed the hindrance of the TX membrane to the diffusion of molecules using simulation. The joint consideration of the hindrance of the TX membrane and the absorbing RX is cumbersome for theoretical analysis. We will investigate the impact of the TX membrane on molecule diffusion in the propagation environment in future work. In Fig. 7(b) and Table \ref{tab31}, we relax this assumption by treating the TX membrane as a reflecting boundary for   molecules in the propagation environment and perform the corresponding particle-based simulation \cite{andrews2004stochastic}. Results indicate that an obvious deviation is caused only when the TX and RX are very close to each other.
	\end{itemize}
	Based on assumptions A2 and A3, if a vesicle hits the TX membrane, it fuses to the membrane with a probability of $k_\mathrm{f}\sqrt{\frac{\pi\Delta t}{D_\mathrm{v}}}$ during a time interval $\Delta t$ \cite{andrews2009accurate}. We define this probability as the MF probability\footnote{Based on this definition, the MF-based TX can be regarded as a TX with a semi-permeable boundary as in \cite{schafer2019spherical}. We clarify that the major difference between our work and \cite{schafer2019spherical} is the method of analysis. The authors in \cite{schafer2019spherical} applied the transfer function approach to investigate the molecule concentration within the $\mathrm{TX}$, while we focus on the end-to-end communication channel and derive analytical expressions for the CIR in this channel.}. The equation of the MF probability is accurate if $k_\mathrm{f}\ll\sqrt{\frac{D_\mathrm{v}}{\pi\Delta t}}$. The MF probability is applied in the simulation process in Section \ref{SF}. After MF, the type $\sigma$ molecules stored by the vesicle are instantaneously released into the propagation environment. The location and time for the occurrence of MF are the initial location and time for molecules to start moving in the propagation environment. We note that the time scale for the MF process is ignored since it is relatively small compared to the entire transmission process \cite{haluska2006time}.
	\subsection{Transceiver Mobility} 
	We consider the following two cases for TX and RX mobility:
	\subsubsection{Static MC}
	We choose the center of the TX as the origin of the environment, as depicted in Fig. 1(a). The center of the RX is distance $l$ away from the center of the TX. We consider an impulse of $N_\mathrm{v}$ vesicles released within the TX at time $t=0$. In this case, the channel response is time-invariant.
	\subsubsection{Mobile MC} 
	We choose the center of the TX at time $t=0$ as the origin of the environment, as depicted in Fig. 1(b). The TX and RX start to diffuse randomly with constant diffusion coefficients $D_{\ss\T}$ and $D_{\ss\R}$, respectively, from time $t=0$, and the distance between their centers at $t=0$ is $l_0$. Different from the static MC case, the channel response is time-variant such that the CIR depends on the release time of vesicles. To ensure generality, we consider an impulse of $N_\mathrm{v}$ vesicles released within the TX at time $t'$. We further denote $l_{t'}$ as the distance between the centers of the TX and RX at time $t'$. In this model, we assume that the diffusion of vesicles is relative to the TX and remains symmetric.
	\subsection{Bit Transmission}
	We assume for both static and mobile scenarios that information transmitted from the TX to the RX is encoded into a sequence of binary bits with length $W$. The sequence of binary bits is $\mathbf{b}_{1:W}=\left[b_1, b_2,...b_W\right]$ and $b_w$ is the $w$th bit. We denote $\phi$ as the bit interval length. We assume that each bit is transmitted at the beginning of the current bit interval with probabilities $\mathrm{Pr}(b_w=1)=P_1$ and $\mathrm{Pr}(b_w=0)=P_0=1-P_1$, where $\mathrm{Pr}(\cdot)$ denotes the probability. For the static scenario, we consider that $b_1$ is transmitted at time $t=0$. For the mobile scenario, we consider that $b_1$ is transmitted at time $t'$ such that the distance between the centers of the TX and RX is a random variable (RV) when $b_1$ is transmitted. We adopt ON/OFF keying for the modulation, which means that $N_\mathrm{v}$ vesicles are released from the TX's center at the beginning of the bit interval to transmit bit $1$ while no vesicle is released to transmit bit 0. We further assume that the TX and RX are perfectly synchronized since several practical synchronization schemes for MC have been studied and proposed in \cite{jamali2017symbol}. For demodulation of $b_w$ at the RX, we adopt a widely-investigated threshold detector that compares the number of molecules absorbed during the $w$th bit interval with a detection threshold \cite{cao2019diffusive}. In this paper, we consider the threshold of the detector is fixed for simplicity. More complex detectors, e.g., an adaptive threshold detector, are summarized in \cite{kuran2020survey}.
	
	\section{Analysis of Channel Impulse Response for Static MC}\label{dCIRs}
	In this section, we first derive the molecule release probability and the fraction of molecules released from the TX membrane due to the impulsive emission of vesicles from the TX's center. We define the molecule release probability as the probability of one molecule being released during the time interval $[t, t+\delta t]$ from the TX membrane, when this molecule is released from the origin at time $t=0$. Here, $\delta t$ represents a very small value of time $t$. We then derive the molecule hitting probability at the RX when the TX releases molecules uniformly over the TX membrane. Using the previously-derived release probability and hitting probability, we finally derive the end-to-end i) molecule hitting probability, ii) fraction of molecules absorbed, and iii) the asymptotic fraction of molecules absorbed at the RX as $t\rightarrow\infty$ due to the impulsive emission of vesicles from the TX's center. We define the molecule hitting probability and end-to-end molecule hitting probability as the probability of one molecule hitting the RX during the time interval $[t, t+\delta t]$ when this molecule is released from the TX membrane and TX's center at time $t=0$, respectively.
	\subsection{Release Probability from TX Membrane}\label{rr}
	As MF guarantees the release of molecules from vesicles to the propagation environment, the probability of releasing molecules equals the vesicle fusion probability. Thus, we need to obtain the distribution of vesicle concentration within the TX to derive the release probability from the TX membrane. In the spherical coordinate system, we denote $C(r,t)$, $0\leq r\leq r_{\ss\T}$, as the distribution of vesicle concentration at time $t$ and distance $r$ from the TX's center. When an impulse of vesicles is released from the TX's center at $t=0$, the initial condition is expressed as \cite[eq. 3(c)]{dincc2018impulse}
	\begin{align}\label{IC}
		C(r,t\rightarrow 0)=\frac{1}{4\pi r^2}\delta(r),
	\end{align}
	where $\delta(\cdot)$ is the Dirac delta function. We then apply Fick's second equation to describe the diffusion of vesicles within the TX. In general, Fick's second equation is a spatially 3D partial differential equation. As we consider that vesicles are released from the TX's center, the TX model is spherically symmetric such that the diffusion of vesicles is described as \cite[eq. (2.7)]{berg1993random}
	\begin{align}\label{fsl}
		D_\mathrm{v}\frac{\partial^2\left(rC(r,t)\right)}{\partial r^2}=\frac{\partial \left(rC(r,t)\right)}{\partial t}.
	\end{align}
	
	Based on assumption A2, the boundary condition is described by the irreversible reaction given by \eqref{IR}, which can be characterized by the radiation boundary condition \cite[eq. (3.25)]{schulten2000lectures}
	\begin{align}\label{bc}
		-D_\mathrm{v}\frac{\partial C(r,t)}{\partial r}\big|_{r=r_{\ss\T}}=k_\mathrm{f}C(r_{\ss\T},t),
	\end{align}
	where the left-hand side represents the diffusion flux\footnote{The diffusion flux $[\mathrm{molecule}\cdot\mathrm{m}^{-2}\cdot\mathrm{s}^{-1}]$ is the rate of movement of molecules across a unit area in unit time \cite{berg1993random}.} over the TX membrane. Since the flux is in the positive radial direction, the right-hand side is positive.
	
	negative sign on the right-hand side indicates that the condition is over the inner boundary.
	
	Based on the initial condition in \eqref{IC}, Fick's second law in \eqref{fsl}, and the boundary condition in \eqref{bc}, we derive the analytical expression for the molecule release probability from the TX membrane, denoted by $f_\mathrm{r}(t)$, in the following theorem:
	\begin{theorem}\label{theorem1}
		The molecule release probability from the TX membrane at time $t$ is given by
		\begin{align}\label{ff}
			f_\mathrm{r}(t)\!=\!\!\sum_{n=1}^{\infty}\frac{4r_{\ss\T}^2k_\mathrm{f}\lambda_n^3}{2\lambda_nr_{\ss\T}\!-\!\mathrm{sin}\left(2\lambda_nr_{\ss\T}\right)}j_0(\lambda_nr_{\ss\T})\exp\!\left(-D_\mathrm{v}\lambda_n^2t\right),
		\end{align}
		where $j_0(\cdot)$ is the zeroth order spherical Bessel function of the first type \cite{olver1960bessel} and $\lambda_n$ is obtained by solving 
		\begin{align}\label{Dvl}
			-D_\mathrm{v}\lambda_nj_0'\left(\lambda_nr_{\ss\T}\right)=k_\mathrm{f}j_0\left(\lambda_nr_{\ss\T}\right),
		\end{align}
		with $j_0'(z)=\frac{\partial j_0(z)}{\partial z}$ and $n=1,2,...$. In particular, \eqref{Dvl} is directly obtained from the radiation boundary condition in \eqref{bc} with eigenfunction $j_0\left(\lambda_nr_{\ss\T}\right)$ and discrete eigenvalue $\lambda_{n}$.
	\end{theorem}
	\begin{IEEEproof}
		Please see Appendix \ref{app}.
	\end{IEEEproof}
	
	\begin{remark}
		We note that \eqref{ff} becomes accurate when $n$ increases. We also note that the number $n$ applied in \eqref{ff} can be determined mathematically. We denote $n^*$ as the maximum $n$ applied in \eqref{ff} and $f_{\mathrm{r},n}(t)=\frac{4r_{\ss\T}^2k_\mathrm{f}\lambda_n^3}{2\lambda_nr_{\ss\T}-\mathrm{sin}\left(2\lambda_nr_{\ss\T}\right)}j_0(\lambda_nr_{\ss\T})\exp\left(-D_\mathrm{v}\lambda_n^2t\right)$. When $n\geq n^*$, $f_{\mathrm{r},n}(t)=0$. The value of $n^*$ increases when $t$ decreases. In this paper, we choose $t=\hat{t}=0.01\;\mathrm{s}$ as a criterion to obtain $n^*$. Therefore, $n^*$ is obatined by numerically finding the minimum $n$ satisfying $f_{\mathrm{r},n}(\hat{t})=0$.
	\end{remark}
	
	We denote $F_\mathrm{r}(t)$ as the fraction of molecules released by time $t$ and obtain it via $F_\mathrm{r}(t)=\int_{0}^{t}f_\mathrm{r}(u)\mathrm{d}u$. We present $F_\mathrm{r}(t)$ in the following corollary:
	\begin{corollary}
		The fraction of molecules released from the TX by time $t$ is given by
		\begin{align}\label{num}
			F_\mathrm{r}(t)=\sum_{n=1}^{\infty}\frac{4r_{\ss\T}^2 k_\mathrm{f}\lambda_nj_0(\lambda_nr_{\ss\T})}{D_\mathrm{v}\left(2\lambda_nr_{\ss\T}-\mathrm{sin}(2\lambda_nr_{\ss\T})\right)}\left(1-\exp\left(-D_\mathrm{v}\lambda_n^2t\right)\right).
		\end{align}
	\end{corollary}
	As each molecule has the same probability to be released by time $t$, i.e., $F_\mathrm{r}(t)$, the number of molecules released by time $t$ is $N_\mathrm{v}\eta F_\mathrm{r}(t)$.
	\subsection{Hitting Probability at RX with Uniform Release of Molecules}\label{thr}
	In this subsection, we derive the hitting probability when molecules are uniformly released from the TX membrane, i.e., ignoring the internal molecules' propagation within vesicles inside the TX and the TX's MF process. This hitting probability establishes the foundation for deriving the end-to-end molecule hitting probability at the RX surface in Section \ref{end}. We consider that molecules are initially uniformly distributed over the TX membrane and released simultaneously at $t=0$. We denote $\Omega_{\ss\T}$ as the membrane area and $p_\mathrm{u}(t)$ as the hitting probability at the RX due to the uniform release of molecules. We clarify that uniformly-distributed molecules means that the likelihood of a molecule released from any point on the TX membrane is the same. We denote this probability by $\rho$, where we have $\rho=\left(4\pi r_{\ss\T}^2\right)^{-1}$. We further consider an arbitrary point $\alpha$ on the TX membrane. Based on \cite[eq. (9)]{heren2015effect}, the hitting probability of a molecule at the RX at time $t$ when the molecule is released from the point $\alpha$ at time $t=0$, $p_\alpha(t)$, is given by
	\begin{align}\label{pa}
		p_\alpha(t)=\frac{r_{\ss\R}(l_\alpha-r_{\ss\R})}{l_\alpha\sqrt{4\pi D_{\sigma}t^3}}\exp\left(-\frac{\left(l_\alpha-r_{\ss\R}\right)^2}{4D_{\sigma}t}-k_\mathrm{d}t\right),
	\end{align}
	where $l_\alpha$ is the distance between the point $\alpha$ and the center of the RX.
	
	Given that molecules are distributed uniformly over the TX membrane, $p_\mathrm{u}(t)$ is obtained by taking the surface integral of $p_\alpha(t)$ over the spherical TX membrane. Using this method, we solve $p_\mathrm{u}(t)$ in the following lemma:
	\begin{lemma}\label{hrb}
		The molecule hitting probability at the RX at time $t$ when the TX uniformly releases molecules over the TX membrane at time $t=0$ is given by
		\begin{align}\label{pbt}
			p_\mathrm{u}(t)\!\!=\!\!\frac{2\rho r_{\ss\T}r_{\ss\R}}{l}\!\sqrt{\frac{\pi D_{\sigma}}{t}}\!\left[\!\exp\!\left(\!\!-\frac{\beta_1}{t}\!-\!k_\mathrm{d}t\!\right)\!-\!\exp\!\left(\!-\frac{\beta_2}{t}\!-\!k_\mathrm{d}t\right)\!\right],
		\end{align}
		where $\beta_1=\frac{(r_{\ss\T}+r_{\ss\R})(r_{\ss\T}+r_{\ss\R}-2l)+l^2}{4D_{\sigma}}$ and $\beta_2=\frac{(r_{\ss\T}-r_{\ss\R})(r_{\ss\T}-r_{\ss\R}+2l)+l^2}{4D_{\sigma}}$.
	\end{lemma}
	\begin{IEEEproof}
		Please see Appendix \ref{ab}.
	\end{IEEEproof}
	
	We denote $P_\mathrm{u}(t)$ as the fraction of molecules absorbed by time $t$ when the TX uniformly releases molecules and obtain it via $P_\mathrm{u}(t)=\int_{0}^{t}p_\mathrm{u}(u)\mathrm{d}u$. We present $P_\mathrm{u}(t)$ in the following corollary:
	\begin{corollary}
		The fraction of molecules absorbed at the RX by time $t$ when the TX uniformly releases molecules over the TX membrane at time $t=0$ is given by \eqref{Pbt} on the top of next page.
		\begin{figure*}[!t]
			\begin{align}\label{Pbt}
				P_\mathrm{u}(t)=&\frac{\rho r_{\ss\T}r_{\ss\R}\pi}{l}\sqrt{\frac{D_{\sigma}}{k_\mathrm{d}}}\left[\exp\left(-2\sqrt{\beta_1 k_\mathrm{d}}\right)\mathrm{erfc}\left(\sqrt{\frac{\beta_1}{t}}-\sqrt{k_\mathrm{d}t}\right)-\exp\left(2\sqrt{\beta_1 k_\mathrm{d}}\right)\mathrm{erfc}\left(\sqrt{\frac{\beta_1}{t}}+\sqrt{k_\mathrm{d}t}\!\right)-\exp\left(-2\sqrt{\beta_2 k_\mathrm{d}}\right)\right.\notag\\&\left.\times\mathrm{erfc}\left(\sqrt{\frac{\beta_2}{t}}-\sqrt{k_\mathrm{d}t}\right)+\exp\left(2\sqrt{\beta_2 k_\mathrm{d}}\right)\mathrm{erfc}\left(\sqrt{\frac{\beta_2}{t}}+\sqrt{k_\mathrm{d}t}\right)\right].
			\end{align}
			\hrulefill \vspace*{-1pt}
		\end{figure*}
	\end{corollary}
	
	\subsection{End-to-End Hitting Probability at the RX}\label{end}
	We denote $p_\mathrm{e}(t)$ as the end-to-end molecule hitting probability at the RX when an impulse of vesicles is released from the TX's center at time $t=0$. The molecule release probability from the TX membrane at time $u$, $0\leq u\leq t$, is given by \eqref{ff}, which is $f_\mathrm{r}(u)$. For molecules released at time $u$, the molecule hitting probability at the RX at time $t$ is given by \eqref{pbt}, which is $p_\mathrm{u}(t-u)$. Based on $f_\mathrm{r}(u)$ and $p_\mathrm{u}(t-u)$, $p_\mathrm{e}(t)$ is given by $	p_\mathrm{e}(t)=\int_{0}^{t}p_\mathrm{u}(t-u)f_\mathrm{r}(u)\mathrm{d}u$. Applying \eqref{ff} and \eqref{pbt} to this equation, we derive $p_\mathrm{e}(t)$ in the following theorem:
	\begin{theorem}
		The end-to-end molecule hitting probability at the RX at time $t$ for an impulsive emission of vesicles from the TX's center at time $t=0$ is given by
		\begin{align}\label{pvt}
			p_\mathrm{e}(t)=&\frac{8\rho r_{\ss\T}^3r_{\ss\R}k_\mathrm{f}\sqrt{\pi D_{\sigma}}\exp\left(-k_\mathrm{d}t\right)}{l}\sum_{n=1}^{\infty}\frac{\lambda_n^3j_0\left(\lambda_nr_{\ss\T}\right)}{2\lambda_nr_{\ss\T}-\mathrm{sin}\left(2\lambda_nr_{\ss\T}\right)}\notag\\&\times\left[\varepsilon(\beta_1,t)-\varepsilon(\beta_2,t)\right],
		\end{align}
		where $\varepsilon(\beta,t)\!\!=\!\!\!\int_{0}^{t}\!\!\frac{1}{\sqrt{t\!-\!u}}\exp\left(-\frac{\beta}{t-u}-(D_\mathrm{v}\lambda_n^2-k_\mathrm{d})u\right)\mathrm{d}u$. $\varepsilon(\beta,t)$ can be calculated numerically using MATLAB or Mathematica, e.g., using the built-in function \textit{integral} in MATLAB or the built-in function \textit{NIntegrate} in Mathematica.
	\end{theorem}
	
	We denote $P_\mathrm{e}(t)$ as the end-to-end fraction of molecules absorbed at the RX by time $t$ and obtain it via $P_\mathrm{e}(t)=\int_{0}^{t}p_\mathrm{e}(u)\mathrm{d}u=\int_{0}^{t}P_\mathrm{u}(t-u)f_\mathrm{r}(u)\mathrm{d}u$. We present $P_\mathrm{e}(t)$ in the following corollary:
	\begin{corollary}
		The end-to-end fraction of molecules absorbed at the RX by time $t$ for an impulsive emission of vesicles from the TX's center at time $t=0$ is given by
		\begin{align}\label{Pv}
			P_\mathrm{e}(t)=&\frac{4r_{\ss\T}^3r_{\ss\R}k_\mathrm{f}\rho\pi}{l}\sqrt{\frac{D_{\sigma}}{k_\mathrm{d}}}\sum_{n=1}^{\infty}\frac{\lambda_n^3j_0(\lambda_nr_{\ss\T})}{2\lambda_nr_{\ss\T}-\mathrm{sin}(2\lambda_nr_{\ss\T})}\left[\epsilon_1(\beta_1,t)\notag\right.\\&\left.-\epsilon_2(\beta_1,t)-\epsilon_1(\beta_2,t)+\epsilon_2(\beta_2,t)\right],
		\end{align}
		where $\epsilon_\varpi(\beta,t)$, $\varpi=1, 2$, is given by
		\begin{align}
			\epsilon_\varpi(\beta,t)&=\int_{0}^{t}\exp\left((-1)^\varpi2\sqrt{\beta k_\mathrm{d}}-D_\mathrm{v}\lambda_n^2u\right)\notag\\&\times\mathrm{erfc}\left(\sqrt{\frac{\beta}{t-u}}+(-1)^\varpi\sqrt{k_\mathrm{d}(t-u)}\right)\mathrm{d}u.
		\end{align} 
	\end{corollary}
	
	We further denote $P_{\mathrm{e}, \infty}$ as the end-to-end asymptotic fraction of molecules absorbed as $t\rightarrow\infty$. We present $P_{\mathrm{e}, \infty}$ in the following corollary:
	\begin{corollary}\label{asy}
		As $t\rightarrow\infty$, the end-to-end asymptotic fraction of molecules absorbed at the RX for an impulsive emission of vesicles from the TX's center at time $t=0$ is given by 
		\begin{align}\label{pvasy}
			P_{\mathrm{e}, \infty}\!=\!\frac{2r_{\ss\T}r_{\ss\R}\rho\pi}{l}\!\sqrt{\frac{D_{\sigma}}{k_\mathrm{d}}}\!\left[\!\exp\!\left(\!-2\sqrt{\beta_1 k_\mathrm{d}}\right)\!-\!\exp\!\left(\!-2\sqrt{\beta_2 k_\mathrm{d}}\right)\right].
		\end{align}
		
		\begin{IEEEproof}
			Please see Appendix \ref{co}.
		\end{IEEEproof}
	\end{corollary}
	\begin{remark}\label{asf}
		Eq. \eqref{pvasy} indicates that the end-to-end asymptotic fraction of molecules absorbed only depends on the size of the TX and RX, the distance between centers of the TX and RX, and the diffusion coefficient and molecule degradation rate in the propagation environment. $P_{\mathrm{e}, \infty}$ is $\emph{independent}$ of other TX properties, e.g., the forward reaction rate in MF and the vesicle diffusion coefficient. This is because, with sufficient time, all molecules are released from the TX.
	\end{remark}
	
	\section{Analysis of Expected Channel Impulse Response for Mobile MC}\label{DCIRm}
	In this section, we consider the scenario that both the TX and RX are mobile using diffusion. Since the distance between the centers of the TX and RX is a RV, the CIR is time-variant and modeled as a stochastic process \cite{ahmadzadeh2018stochastic}. Hence, in this section we focus on the expected CIR, i.e., the mean of the CIR over the varying distance. We first derive the expected molecule hitting probability at the RX when the TX releases molecules uniformly over the TX membrane. Using this probability, we then derive the expected end-to-end molecule hitting probability at the RX due to an impulsive emission of vesicles from the TX's center. For the mobile scenario, we define the expected hitting probability and expected end-to-end hitting probability as the probability of one molecule hitting the RX during the time interval $[t'+\tau, t'+\tau+\delta\tau]$, $\tau\geq 0$, when this molecule is released from the TX membrane and TX's center at time $t'$, respectively. Here, $\tau$ is a relative time of observing absorbed molecules at the RX for a fixed $t'$ and $\delta\tau$ represents a very small value of time $\tau$.
	\subsection{Expected Hitting Probability at Mobile RX with Uniform Release of Molecules}
	We denote $g(l_{t'}|l_0)$ as the probability density function (PDF) of $l_{t'}$, which is the distance between centers of the TX and RX at time $t'$, given that the initial distance between the centers of the TX and RX is $l_0$ at $t=0$. The coordinates of the TX's center and RX's center at time $t'$ are $\left[x_{\ss\T}(t'), y_{\ss\T}(t'), z_{\ss\T}(t')\right]$ and $\left[x_{\ss\R}(t'), y_{\ss\R}(t'), z_{\ss\R}(t')\right]$, respectively. As both TX and RX perform Brownian motion, the displacement of each coordinate follows a Gaussian distribution. $l_{t'}$ is calculated as $l_{t'   }=\sqrt{(x_{\ss\T}(t')-x_{\ss\R}(t'))^2+(y_{\ss\T}(t')-y_{\ss\R}(t'))^2+(z_{\ss\T}(t')-z_{\ss\R}(t'))^2}$, where $x_{\ss\T}(t')-x_{\ss\R}(t')$, $y_{\ss\T}(t')-y_{\ss\R}(t')$, and $z_{\ss\T}(t')-z_{\ss\R}(t')$ follow the Gaussian distribution. Thus, $\frac{l_{t'}}{\sqrt{2D_1t'}}$ follows a noncentral chi distribution with three degrees of freedom \cite{Miller1958Generalized}, and $g(l_{t'}|l_0)$ is given by \cite[eq. (8)]{cao2019diffusive}
	\begin{align}\label{dst}
		g(l_{t'}|l_0)=\frac{l_{t'}^\frac{3}{2}}{2D_1t'\sqrt{l_0}}\exp\left(-\frac{l_{t'}^2+l_0^2}{4D_1t'}\right)I_\frac{1}{2}\left(\frac{l_{t'}l_0}{2D_1t'}\right),
	\end{align}
	where $D_1=D_{\ss\T}+D_{\ss\R}$ and $I_\frac{1}{2}(\cdot)$ is the modified Bessel function of the first kind \cite{schulten2000lectures}. We note that $D_{\ss\T}$, $D_{\ss\R}$, and $t'$ should not be very large. Otherwise, $l_{t'}$ may become less than $r_{\ss\T}+r_{\ss\R}$ such that the TX and RX overlap with each other.
	
	We denote $p_\mathrm{u,m}(\tau,t')$ as the $\emph{expected}$ hitting probability at the mobile RX at time $t'+\tau$ when the mobile TX releases molecules uniformly over the TX membrane at time $t'$. To describe the relative motion between released molecules and the RX, we define an effective diffusion coefficient $D_2$ as $D_2=D_{\ss\R}+D_{\sigma}$ \cite{2017Diffusive}. By replacing $l$, $t$, and $D_{\sigma}$ in \eqref{pbt} with $l_{t'}$, $\tau$, and $D_2$, respectively, we obtain the hitting probability at time $t'+\tau$, i.e., $p_\mathrm{u}(\tau)\big|_{l=l_{t'}, D_\sigma=D_2}$, for a given $l_{t'}$. We note that $l_{t'}$ is a RV with the PDF given by \eqref{dst}. Hence, we obtain $p_\mathrm{u,m}(\tau,t')$ as $p_\mathrm{u,m}(\tau,t')=\int_{0}^{\infty}g(l_{t'}|l_0)p_\mathrm{u}(\tau)\big|_{l=l_{t'}, D_\sigma=D_2}\mathrm{d}l_{t'}$ and present it in the following lemma:
	\begin{lemma}
		The expected molecule hitting probability at the mobile RX at time $t'+\tau$ when the mobile TX uniformly releases the molecules over the TX membrane at time $t$ is given by \eqref{eht} on the top of next page,
		\begin{figure*}[!t]
			\begin{align}\label{eht}
				&p_\mathrm{u,m}(\tau,t')=\frac{\rho r_{\ss\T}r_{\ss\R}D_2}{l_0}\sqrt{\frac{\pi}{D_1t'+D_2\tau}}\exp\left(-\frac{l_0^2}{4(D_1t'+D_2\tau)}\right)\bigg\{\exp\left(-\frac{(r_{\ss\T}+r_{\ss\R})^2}{4(D_1t'+D_2\tau)}\right)\bigg[\theta(r_{\ss\T},\tau,t')\theta(r_{\ss\R},\tau,t')\notag\\&\times\mathrm{erfc}\left(-\hat{\vartheta}(r_{\ss\T},\tau,t')-\hat{\vartheta}(r_{\ss\R},\tau,t')-\vartheta(\tau,t')\right)-\theta(-r_{\ss\T},\tau,t')\theta(-r_{\ss\R},\tau,t')\times\mathrm{erfc}\left(\vartheta(\tau,t')-\hat{\vartheta}(r_{\ss\T},\tau,t')-\hat{\vartheta}(r_{\ss\R},\tau,t')\right)\bigg]\notag\\&-\exp\left(-\frac{(r_{\ss\T}-r_{\ss\R})^2}{4(D_1t'+D_2\tau)}\right)\left[\theta(r_{\ss\R},\tau,t')\theta(-r_{\ss\T},\tau,t')\mathrm{erfc}\left(-\hat{\vartheta}(r_{\ss\R},\tau,t')+\hat{\vartheta}(r_{\ss\T},\tau,t')-\vartheta(\tau,t')\right)-\theta(-r_{\ss\R},\tau,t')\theta(r_{\ss\T},\tau,t')\notag\right.\\&\left.\times\mathrm{erfc}\left(\vartheta(\tau,t')-\hat{\vartheta}(r_{\ss\R},\tau,t')+\hat{\vartheta}(r_{\ss\T},\tau,t')\right)\right]\bigg\},
			\end{align}
			\hrulefill \vspace*{-10pt}
		\end{figure*}
		where $\theta(\zeta,\tau,t')=\exp\left(\frac{l_0\zeta}{2(D_1t'+D_2\tau)}\right)$, $\hat{\vartheta}(\zeta,\tau,t')=\frac{\zeta}{2}\sqrt{\frac{D_1t'}{D_2\tau(D_1t'+D_2\tau)}}$, and $\vartheta(\tau,t')=\frac{l_0}{2}\sqrt{\frac{D_2\tau}{D_1t'(D_1t'+D_2\tau)}}$.
	\end{lemma}
	
	\subsection{Expected End-to-End Hitting Probability at the Mobile RX}
	We denote $p_\mathrm{e,m}(\tau,t')$ as the \emph{expected} end-to-end molecule hitting probability at the mobile RX at time $t'+\tau$ when an impulse of vesicles is released from the TX's center at time $t'$. The molecule release probability from the TX at time $t'+u$, $0\leq u\leq\tau$, is given by \eqref{ff}, which is $f_\mathrm{r}(u)$. For molecules released at time $t'+u$, the expected hitting probability at time $t'+\tau$ at the RX is given by \eqref{eht}, which is $p_\mathrm{u,m}(\tau-u,t'+u)$. Based on that, $p_\mathrm{e,m}(\tau,t')$ is given by $p_\mathrm{e,m}(\tau,t')=\int_{0}^{\tau}f_\mathrm{r}(u)p_\mathrm{u,m}(\tau-u, t'+u)\mathrm{d}u$. Applying \eqref{ff} and \eqref{eht} to this equation, we derive $p_\mathrm{e,m}(\tau, t')$ in the following theorem:
	\begin{theorem}
		The expected end-to-end molecule hitting probability at the mobile RX at time $t'+\tau$ for an impulsive emission of vesicles from the mobile TX's center at time $t'$ is given by
		\begin{align}\label{pvm}
			&p_\mathrm{e,m}(\tau, t')=\sum_{n=1}^{\infty}\frac{4r_{\ss\T}^3r_{\ss\R}\rho D_2k_\mathrm{f}\lambda_n^3\sqrt{\pi}j_0(\lambda_nr_{\ss\T})}{l_0(2\lambda_nr_{\ss\T}-\mathrm{sin}(2\lambda_nr_{\ss\T}))}\notag\\&\times\left[\kappa\left((r_{\ss\T}+r_{\ss\R})^2-2l_0r_{\ss\T}-2l_0r_{\ss\R}, r_{\ss\T}+r_{\ss\R},-1,\tau,t'\right)\right.\notag\\&\left.-\kappa((r_{\ss\T}+r_{\ss\R})^2+2l_0r_{\ss\T}+2l_0r_{\ss\R}, r_{\ss\T}+r_{\ss\R},1,\tau,t')\right.\notag\\&\left.-\kappa((r_{\ss\T}-r_{\ss\R})^2-2l_0r_{\ss\R}+2l_0r_{\ss\T}, r_{\ss\R}-r_{\ss\T},-1,\tau,t')\right.\notag\\&\left.+\kappa((r_{\ss\T}-r_{\ss\R})^2+2l_0r_{\ss\R}-2l_0r_{\ss\T}, r_{\ss\R}-r_{\ss\T},1,\tau,t')\right],
		\end{align}
		where $\kappa(\zeta_1, \zeta_2, \zeta_3, \tau, t')$ is given by \eqref{nu} on the top of next page with $\mu(\tau,t')=D_1t'+D_2\tau$ and $\nu=D_1-D_2$.
		\begin{figure*}[!t]
			\begin{align}\label{nu}
				&\kappa(\zeta_1, \zeta_2, \zeta_3, \tau, t')=\int_{0}^{\tau}\frac{1}{\sqrt{\mu(\tau,t')+\nu u}}\exp\left(-\frac{l_0^2+\zeta_1}{4(\mu(\tau,t')+\nu u)}-k_\mathrm{d}\tau+\left(k_\mathrm{d}-D_\mathrm{v}\lambda_n^2\right)u\right)\notag\\&\times\mathrm{erfc}\left(\frac{\zeta_3 l_0}{2}\sqrt{\frac{D_2(\tau-u)}{D_1(t'+u)(\mu(\tau,t')+\nu u)}}-\frac{\zeta_2}{2}\sqrt{\frac{D_1(t'+u)}{D_2(\tau-u)(\mu(\tau,t')-\nu u)}}\right)\mathrm{d}u,
			\end{align}
			\hrulefill \vspace*{-10pt}
		\end{figure*}
	\end{theorem}
	
	\begin{remark}
		The expected end-to-end fraction of molecules absorbed by time $t'+\tau$ at the mobile RX for an impulsive emission of vesicles from the mobile TX's center at time $t'$ is obtained from the integral of \eqref{pvm}, wherein the double integral can be calculated using MATLAB or Mathematica.
	\end{remark}
	
	\section{Error Probability for Static and Mobile MC}\label{sber}
	In this section, we consider the transmission of a sequence of bits from the TX to the RX. For the static scenario, we calculate the average BER, following the formulation in \cite{ahmadzadeh2018stochastic} and similar studies. For the mobile scenario, we first calculate the PMF for the number of molecules absorbed  within a bit interval. Based on the derived PMF, we then calculate the average BER in the mobile scenario. 
	\subsection{Average BER for Static MC}\label{absm}
	Due to the delay of molecule transport, the RX may receive molecules released from the current and all previous bit intervals. According to \eqref{Pv}, the fraction of molecules absorbed during the $w$th bit interval when vesicles for storing those molecules released at the beginning of the $i$th bit interval, $0\leq i\leq w$, is expressed as $P_\mathrm{e}((w-i+1)\phi)-P_\mathrm{e}((w-i)\phi)$, where $\phi$ is the bit interval length. We denote $N_{w,i}$ as the number of molecules absorbed during the $w$th bit interval for transmitting $b_i$. As the diffusion of vesicles and released molecules are independent and molecules have the same probability to be absorbed, $N_{w,i}$ can be approximated as a Poisson RV when the number of emitted vesicles is large and the success probability $P_\mathrm{e}(t)$ is small \cite{le1960approximation}, i.e., $N_{w,i}\sim b_i\mathrm{Poiss}\left(N_\mathrm{v}\eta\left(P_\mathrm{e}((w-i+1)\phi)-P_\mathrm{e}((w-i)\phi)\right)\right)$. Here, $\mathrm{Poiss}(\cdot)$ refers to a Poisson distribution. This is because sufficient diversity in the vesicle fusion points over the TX membrane is required to match the derivation of the end-to-end channel that includes integration over the entire TX membrane. As the sum of independent Poisson RVs is also a Poisson RV, we model the total number of molecules absorbed during the $w$th bit interval, denoted by $N_w$, as
	\begin{align}\label{pp}
		N_w\sim \mathrm{Poiss}(\psi),
	\end{align}
	where $\psi=N_\mathrm{v}\eta\sum_{i=1}^{w}b_i\left(P_\mathrm{e}((w-i+1)\phi)-P_\mathrm{e}((w-i)\phi)\right)$. Based on \eqref{pp}, the cumulative distribution function (CDF) of the Poisson RV $N_w$ is written as
	\begin{align}\label{pp2}
		\mathrm{Pr}\left(N_w<\xi|\mathbf{b}_{1:w}\right)=\sum_{h=0}^{\xi-1}\frac{\psi^h\exp(-\psi)}{h!}.
	\end{align}
	
	\begin{table*}[!t]
		\centering
		\caption{$R^2$ for the CDF of $N_w$, where the total number of molecules emitted is $N_\mathrm{v}\eta=1000$}\label{tab3}
		{\begin{tabular}{|c|c|c|c|c|c|c|c|}
				\hline
				$N_\mathrm{v}, \eta$&10, 100&20, 50&50, 20&100, 10&200, 5&500, 2&1000, 1\\
				\hline
				$R^2$&0.9771&0.9913&0.9969&0.9962&0.9974&0.9957&0.9974\\
				\hline
			\end{tabular}
		}
	\end{table*}
	
	In Table \ref{tab3}, we calculate the coefficient of determination \cite{lu2016effect}, denoted by $R^2$, for the CDF of $N_w$, where we vary $N_\mathrm{v}$ and $\eta$ and keep the total number of molecules emitted as 1000. We set $w=1$, $D_\mathrm{v}=18\;\mu\mathrm{m}^2/\mathrm{s}$, $k_\mathrm{f}=30\;\mu\mathrm{m}/\mathrm{s}$, and $b_w=1$. For other parameter values, please see Table \ref{tabl}. For the simulation details, please see Section \ref{SF}. The $R^2$ is used to measure the preciseness between simulation and theoretical analysis. Specifically, $R^2$ is given by $R^2=1-SS_\mathrm{res}/SS_\mathrm{tot}$, where $SS_\mathrm{res}$ is the sum of squared differences between the theoretical analysis and simulation, and $SS_\mathrm{tot}$ is the sum of squared differences between the simulation and its mean. Since $SS_\mathrm{res}/SS_\mathrm{tot}$ represents the fraction of variance unexplained, the closer the value of $R^2$ is to 1, the more accurate the theoretical analysis. In this table, we observe that the $R^2$ increases with an increase in the number of vesicles from 10 to 50, which indicates that the accuracy of \eqref{pp2} increases when the number of vesicles emitted increases. This is due to the fact that the Poisson approximation is impacted by two sets of trials, i.e., the number of emitted vesicles and molecules, and the approximation is more sensitive to the number of emitted vesicles since it determines the initialization of the emission of molecules. When $N_\mathrm{v}$ is larger than 50, we observe a small variation of $R^2$ for $N_\mathrm{v}$ from 50 to 1000. We further increase the number of realizations applied in Table \ref{tab3} by ten times for the simulation when $N_\mathrm{v}=10$ and $\eta=100$, and obtain that $R^2=0.9771$. This indicates that the number of realizations applied for the simulation in Table \ref{tab3} is sufficient.
	
	We adopt a simple threshold detector, where $N_w$ is compared with a detection threshold to demodulate $b_w$. We present the threshold detector as
	\begin{align}\label{bk}
		\hat{b}_w=\left\{\begin{array}{lr}
			1, ~~\mathrm{if}~N_w\geq\xi\\
			0, ~~\mathrm{if}~N_w<\xi,
		\end{array}
		\right.
	\end{align}
	where $\hat{b}_w$ is the demodulated bit of $b_w$ at the RX and $\xi$ is the detection threshold. Based on \eqref{pp2} and \eqref{bk}, the error probability of $\hat{b}_w$ for $b_w
	=1$ and $b_w=0$ are
	\begin{align}
		\mathrm{Pr}\left(\hat{b}_w=0|b_w=1, \mathbf{b}_{1:w-1}\right)=\mathrm{Pr}\left(N_w<\xi|b_w=1, \mathbf{b}_{1:w-1}\right)
	\end{align}
	and
	\begin{align}
		\mathrm{Pr}\!\left(\!\hat{b}_w=1|b_w=0, \mathbf{b}_{1:w-1}\!\right)\!=\!1\!-\!\mathrm{Pr}\!\left(\!N_w\!<\!\xi|b_w\!=\!0, \mathbf{b}_{1:w-1}\!\right),
	\end{align}
	respectively. The BER of the $w$th bit, denoted by $Q_\s[w|\mathbf{b}_{1:w-1}]$, is given by
	\begin{align}
		Q_\s[w|\mathbf{b}_{1:w-1}]=&P_1\mathrm{Pr}\left(\hat{b}_w=0|b_w=1, \mathbf{b}_{1:w-1}\right)\notag\\&+P_0\mathrm{Pr}\left(\hat{b}_w=1|b_w=0, \mathbf{b}_{1:w-1}\right).
	\end{align}
	
	We denote $Q_\s$ as the average BER over all realizations of $\mathbf{b}_{1:w-1}$ and all bits from $1$ to $W$ for the static scenario. We present $Q_\s$ as
	\begin{align}
		Q_\s=\frac{1}{W}\sum_{w=1}^{W}\frac{\sum_{\mathbf{b}_{1:w-1}\in\Psi_w}Q_\s[w|\mathbf{b}_{1:w-1}]}{2^{w-1}},
	\end{align}
	where $\Psi_w$ stands for all realizations of $\mathbf{b}_{1:w-1}$.
	
	\subsection{Average BER for Mobile MC}
	\subsubsection{Statistical Analysis of Absorbed Molecules}
	\begin{figure}[!t]
		\begin{center}
			\includegraphics[height=2.2in,width=0.86\columnwidth]{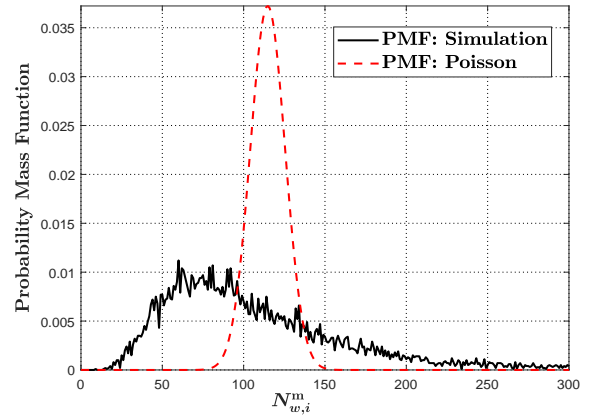}
			\caption{PMF of $N_{w,i}^\mathrm{m}$ plotted by simulation and Poisson distribution,where $w=i=1$, $D_\mathrm{v}=9\;\mu\mathrm{m}^2/\mathrm{s}$, $k_\mathrm{f}=30\;\mu\mathrm{m}/\mathrm{s}$, and $D_{\ss\mathrm{T}}=D_{\ss\mathrm{R}}=8\;\mu\mathrm{m}^2/\mathrm{s}$. For other parameter values, please see Table \ref{tabl}. For the simulation details, please see Section \ref{SF}. }\label{pmfp}\vspace{-0.5em}
		\end{center}
		\vspace{-4mm}
	\end{figure}
	
	We denote $N_{w,i}^\mathrm{m}$ as the number of molecules absorbed during the $w$th bit interval when vesicles that store these molecules are released at the beginning of the $i$th bit interval in the mobile scenario. Due to the distance between the centers of the TX and RX being a RV when releasing molecules, we clarify that $N_{w,i}^\mathrm{m}$ cannot be approximated as a Poisson RV as shown in Fig. \ref{pmfp}. To obtain the accurate PMF for mobile MC, we divide the molecule release process into multiple intervals, detailed as follows.
	
	\begin{figure}[!t]
		\centering	
		\subfigure[Definition of the release duration]{
			\begin{minipage}[t]{1\linewidth}
				\centering
				\includegraphics[height=1in,width=0.95\columnwidth]{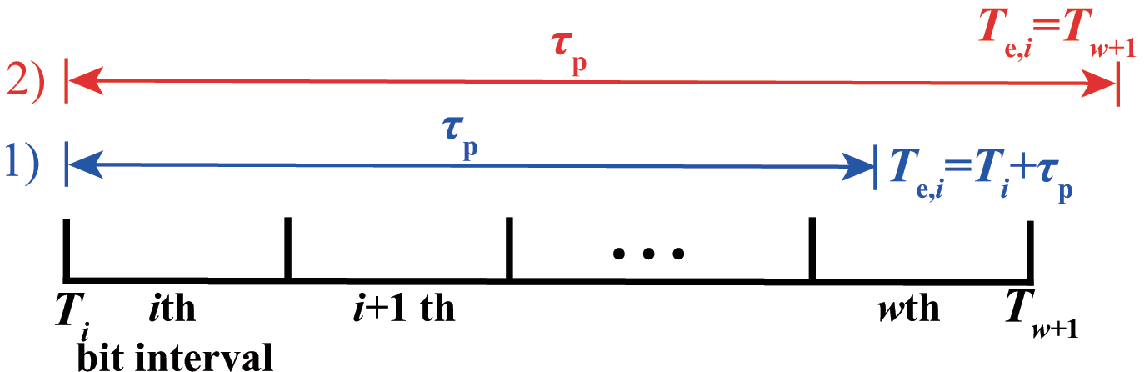}
			\end{minipage}%
		}%
		\quad
		\subfigure[Division of the release duration]{
			\begin{minipage}[t]{1\linewidth}
				\centering
				\includegraphics[height=1.3in,width=0.95\columnwidth]{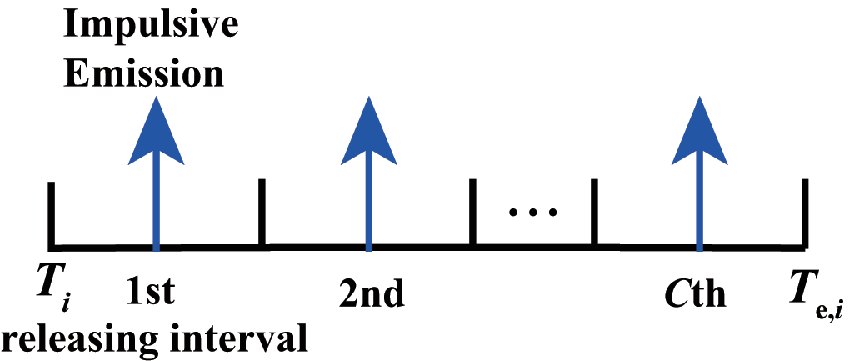}
			\end{minipage}%
		}
		\centering
		\caption{(a). Definition of the release duration $[T_i, T_{\mathrm{e},i}]$, where 1) $T_{\mathrm{e},i}=T_i+\tau_\mathrm{p}$ if  $T_i+\tau_\mathrm{p}\leq T_{w+1}$ and 2) $T_{\mathrm{e},i}=T_{w+1}$ if $T_i+\tau_\mathrm{p}\geq T_{w+1}$.
			(b). Dividing release duration $[T_i, T_{\mathrm{e}, i}]$ into $C$ release intervals, where the release of molecules during the $c$th release interval is assumed as an impulsive emission at time $\hat{\tau}_{i,c}=T_i+\frac{(2c-1)\Delta\tau}{2}$. }\label{tt}
	\end{figure}
	
	To obtain the PMF of $N_{w,i}^\mathrm{m}$, we first define a release duration, denoted by $[T_i, T_{\mathrm{e},i}]$, of transmitting $b_i$ if $b_i=1$ as shown in Fig. 3(a). We denote $T_i$ as the start time of the $i$th bit interval, i.e., $T_i=t'+(i-1)\phi$, and $T_{\mathrm{e},i}$ as the end time of the release duration. We then denote $\tau_\mathrm{p}$ as the period for all molecules released from the TX membrane, where $\tau_\mathrm{p}$ is calculated by finding the minimum $\tau_\mathrm{p}$ that satisfies $F_\mathrm{r}(\tau_\mathrm{p})\geq0.998$ when $n=10000$ in \eqref{num}. Here, we set the threshold at 0.998 such that the calculation of $\tau_\mathrm{p}$ is sufficiently accurate. Since detection in the $w$th bit interval only depends on molecules released before or within the current bit interval, we have $T_{\mathrm{e},i}=T_i+\tau_\mathrm{p}$ if $T_i+\tau_\mathrm{p}\leq T_{w+1}$. Otherwise, $T_{\mathrm{e}, i}=T_{w+1}$. We then consider the discretization of the release duration into $C$ release intervals, as shown in Fig. 3(b). We denote $\Delta\tau$ as the length of each release interval, which is given by $\Delta\tau=\frac{T_{\mathrm{e}, i}-T_i}{C}$. The expected number of molecules released within the $c$th interval is $N_\mathrm{v}\eta (F_\mathrm{r}(T_i+c\Delta\tau)-F_\mathrm{r}(T_i+(c-1)\Delta\tau))$. We denote $\hat{\tau}_{i,c}$ as the middle time of the $c$th release interval, which is obtained as $\hat{\tau}_{i,c}=T_i+\frac{(2c-1)\Delta\tau}{2}$. Since it is cumbersome to incorporate into the theoretical analysis of the precise distance between the TX and RX when each molecule is released, we assume that all molecules released within the $c$th release interval are combined into an impulsive emisson in the middle of the $c$th release interval, i.e., the number of molecules $N_\mathrm{v}\eta (F_\mathrm{r}(T_i+c\Delta\tau)-F_\mathrm{r}(T_i+(c-1)\Delta\tau))$ are impulsively released at time $\hat{\tau}_{i,c}$ from the TX membrane in the subsequent analysis.
	
	We denote $l_{\hat{\tau}_{i,c}}$ as the distance between centers of the TX and RX at time $\hat{\tau}_{i,c}$. We further denote $\mathbf{l}_i$ as the vector that contains all $l_{\hat{\tau}_{i,c}}$ for $c$ from 1 to $C$, i.e., $\mathbf{l}_i=[l_{\hat{\tau}_{i,1}}, l_{\hat{\tau}_{i,2}},...,l_{\hat{\tau}_{i,c}},...,l_{\hat{\tau}_{i,C}}]$. For a given $\mathbf{l}_i$, the expected number of molecules absorbed during the $w$th bit interval for transmitting $b_i$ is
	\begin{align}
		&\psi^\mathrm{m}_{w,i}=b_i\sum_{c=1}^{C}N_\mathrm{v}\eta \left(F_\mathrm{r}(T_i+c\Delta\tau)-F_\mathrm{r}(T_i+(c-1)\Delta\tau)\right)\notag\\&\times\left(\!P_\mathrm{u}(T_{w+1}-\hat{\tau}_{i,c})-P_\mathrm{u}( T_w-\hat{\tau}_{i,c})\right)\big|_{l=l_{\hat{\tau}_{i,c}}, D_\sigma=D_2}.
	\end{align}
	
	For a given $\mathbf{l}_i$, we can approximate $N^\mathrm{m}_{w,i}$ as a Poisson RV that is given by
	\begin{align}\label{po}
		\mathrm{Pr}(N^\mathrm{m}_{w,i}=\xi|\mathbf{l}_i)=\frac{\left(\psi_{w,i}^\mathrm{m}\right)^\xi\exp\left(-\psi_{w,i}^\mathrm{m}\right)}{\xi!}.
	\end{align}
	
	For the mobile TX and RX, each element in $\mathbf{l}_i$ is a RV, and we need to determine the joint PDF for each element in $\mathbf{l}_i$, denoted by $g(\mathbf{l}_i|l_0)$. By considering free diffusion as a memoryless process, i.e., the current state only depends on   the previous one state, $g(\mathbf{l}_i|l_0)$ is given by \cite[eq. (65)]{huang2020initial}
	\begin{align}\label{jd}
		g(\mathbf{l}_i|l_0)=\prod_{c=1}^{C}g(l_{\hat{\tau}_{i,c}}|l_{\hat{\tau}_{i,c-1}}),
	\end{align}
	where $l_{\hat{\tau}_{i,0}}=l_0$.
	
	Based on \eqref{po} and \eqref{jd}, we derive the PMF of $N^\mathrm{m}_{w,i}$ as 
	\begin{align}
		\mathrm{Pr}(\!N^\mathrm{m}_{w,i}\!=\!\xi)\!=\!\!\!\int_{0}^{\infty}\!\!\!...\!\int_{0}^{\infty}\!\!\mathrm{Pr}\!\left(\!N^\mathrm{m}_{w,i}\!=\!\xi|\mathbf{l}_i\right)\!g(\mathbf{l}_i|l_0)\mathrm{d}l_{\hat{\tau}_{i,1}}...\mathrm{d}l_{\hat{\tau}_{i,C}}.
	\end{align}
	We present $\mathrm{Pr}(N^\mathrm{m}_{w,i}=\xi)$ in the following theorem:
	\begin{theorem}
		The PMF for the number of molecules absorbed within the $w$th bit interval when vesicles that store these molecules are released at the beginning of the $i$th bit interval, is given by
		\begin{align}\label{t6}
			\mathrm{Pr}(N^\mathrm{m}_{w,i}=\xi)=&\frac{1}{\xi!}\int_{0}^{\infty}...\int_{0}^{\infty}\left(\psi_{w,i}^\mathrm{m}\right)^\xi\exp\left(-\psi_{w,i}^\mathrm{m}\right)\notag\\&\times\prod_{c=1}^{C}g(l_{\hat{\tau}_{i,c}}|l_{\hat{\tau}_{i,c-1}})\mathrm{d}l_{\hat{\tau}_{i,1}}...\mathrm{d}l_{\hat{\tau}_{i,C}},
		\end{align}
		where the multiple integral can be calculated numerically using the built-in function \textit{NIntegrate} in Mathematica.
	\end{theorem}
	
	\begin{figure}[!t]
		\begin{center}
			\includegraphics[height=2.2in,width=0.86\columnwidth]{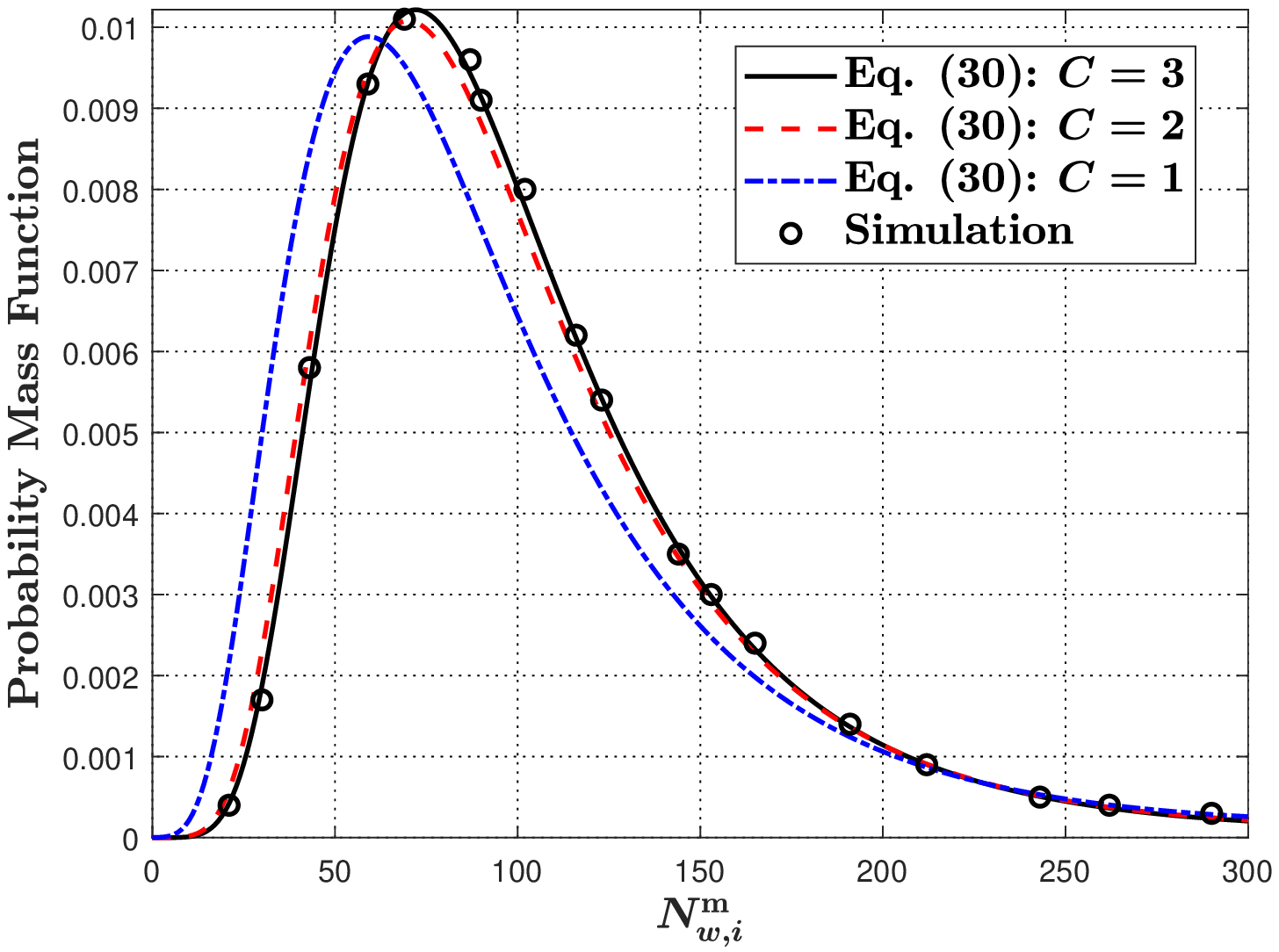}
			\caption{PMF of $N_{w,i}^\mathrm{m}$, where $D_\mathrm{v}=50\;\mathrm{m}^2/\mathrm{s}$ $w=i=1$, $k_\mathrm{f}=30\;\mu\mathrm{m}/\mathrm{s}$, and $D_{\ss\mathrm{T}}=D_{\ss\mathrm{R}}=8\;\mu\mathrm{m}^2/\mathrm{s}$. For other parameter values, please see Table \ref{tabl}. For the simulation details, please see Section \ref{SF}. }\label{num_pmf}\vspace{-0.5em}
		\end{center}
		\vspace{-4mm}
	\end{figure}
	
	In Fig. \ref{num_pmf}, we plot the PMF of $N_{w,i}^\mathrm{m}$ using \eqref{t6} and simulations to validate \eqref{t6}, where the number of intervals from 1 to 3 is adopted. We also set the length of the release duration to be equal to the bit interval $\phi$, i.e, $T_{\mathrm{e},i}-T_i=\phi$.
	We further calculate the R-squared coefficient for \eqref{t6} with varying $C$. The R-squared coefficient is 0.9741, 0.9709, and 0.8637 for $C$ varying from 3 to 1. This indicates that the accuracy of \eqref{t6} increases with an increase in the number of release intervals.
	
	\subsubsection{Average BER}
	We denote $N_w^\mathrm{m}$ as the number of molecules absorbed within the $w$th bit interval, which is given by $N^\mathrm{m}_w=\sum_{i=1}^{w}N^\mathrm{m}_{w,i}$. As in \cite{akdeniz2020equilibrium}, we assume that $\phi$ is sufficiently long such that elements in $\mathbf{N}_w^\mathrm{m}=[N^\mathrm{m}_{w,1}, N^\mathrm{m}_{w,2},...,N^\mathrm{m}_{w,w}]$ are independent of each other. Therefore, the PMF of $N_w^\mathrm{m}$ equals the convolution of the PMF of each element in $\mathbf{N}_w^\mathrm{m}$ \cite{petrov2012sums}. We consider a vector $\mathbf{\hat{i}}=[\hat{i}_1, \hat{i}_2,...,\hat{i}_\Gamma]$, where $\hat{i}_1<\hat{i}_2<...<\hat{i}_\Gamma$, as a subset of vector $[1,2,...,w]$, where $\mathbf{\hat{i}}$ contains all subscripts of bit 1 for a given bit sequence $\mathbf{b}_{1:w}$, i.e., $b_{\hat{i}_1}=b_{\hat{i}_2}=...=b_{\hat{i}_\Gamma}=1$. Thus, we express the PMF of $N^\mathrm{m}_w$ as
	\begin{align}\label{pdf}
		\mathrm{Pr}\left(N_w^\mathrm{m}=\xi|\mathbf{b}_{1:w}\right)\!=&\!\!\left\{\!\mathrm{Pr}\left(\!N_{w,\hat{i}_1}^\mathrm{m}\!=\!0:\xi\!\right)*\mathrm{Pr}\left(N_{w,\hat{i}_2}^\mathrm{m}=0:\xi\right)\right.\notag\\&\left.*...*\mathrm{Pr}\left(N_{w,\hat{i}_\Gamma}^\mathrm{m}=0:\xi\right)\right\}[\xi+1],
	\end{align}
	where $\mathrm{Pr}\left(N_{w,\hat{i}_\gamma}^\mathrm{m}=0:\xi\right)=\left[\mathrm{Pr}\left(N_{w,\hat{i}_\gamma}^\mathrm{m}=0\right), \mathrm{Pr}\left(N_{w,\hat{i}_\gamma}^\mathrm{m}=1\right),...,\mathrm{Pr}\left(N_{w,\hat{i}_\gamma}^\mathrm{m}=\xi\right)\right]$ with each element given by \eqref{t6} and $\left\{\cdot\right\}[\xi+1]$ represents the $(\xi+1)$th element in $\left\{\cdot\right\}$.
	The CDF of $N_w^\mathrm{m}$ is given as $	\mathrm{Pr}\left(N_{w}^\mathrm{m}<\xi|\mathbf{b}_{1:w}\right)=\sum_{h=0}^{\xi-1}\mathrm{Pr}\left(N_w^\mathrm{m}=h|\mathbf{b}_{1:w}\right).$
	
	\begin{table}[!t]
		\centering
		\caption{$R^2$ for the CDF of $N_w^\mathrm{m}$, where $N_\mathrm{v}=200$}\label{tab2}
		{\begin{tabular}{|c|c|c|c|c|c|c|}
				\hline
				$\eta$&5&10&20&30&40&50\\
				\hline
				$R^2$&0.999&0.9988&0.9967&0.9965&0.9503&0.8415\\
				\hline
			\end{tabular}
		}
	\end{table}
	
	In Table \ref{tab2}, we calculate the $R^2$ for the CDF of $N_w^\mathrm{m}$, where we vary $\eta$ and set $N_\mathrm{v}=200$, $w=1$, $D_\mathrm{v}=50\;\mu\mathrm{m}^2/\mathrm{s}$, $k_\mathrm{f}=30\;\mu\mathrm{m}/\mathrm{s}$, $D_{\ss\mathrm{T}}=D_{\ss\mathrm{R}}=8\;\mu\mathrm{m}^2/\mathrm{s}$, and $C=3$. For other parameter values, please see Table \ref{tabl}. We observe that the $R^2$ decreases with an increase in $\eta$ when $N_\mathrm{v}$ is fixed, which indicates that the accuracy of the theoretical analysis for the CDF of $N_w^\mathrm{m}$ decreases with the increase in the number of stored molecules in each vesicle. This is caused by the assumption that all molecules released within the release interval are combined into an impulsive emission. When $\eta$ increases, the number of molecules in each release interval increases such that the inaccuracy of this assumption increases. One method to reduce this inaccuracy is increasing the number of release intervals, which incurs an increase in computational complexity.
	
	By adopting the same detection criterion as in \eqref{bk}, except replacing $N_w$ with $N_w^\mathrm{m}$, we derive the error probability of $\hat{b}_w$ for $b_w=1$ and $b_w=0$ for the mobile scenario, denoted by $\mathrm{Pr_m}\!\left(\!\hat{b}_w\!=\!0|b_w\!=\!1,\!\mathbf{b}_{1:w-1}\!\right)$ and $\mathrm{Pr_m}\left(\hat{b}_w=1|\hat{b}_w=0, \mathbf{b}_{1:w-1}\right)$, as
	\begin{align}\label{prm1}
		\mathrm{Pr_m}\!\left(\!\hat{b}_w\!=\!0|b_w\!=\!1, \mathbf{b}_{1:w-1}\!\right)\!=\mathrm{Pr}\left(N_w^\mathrm{m}<\xi|b_w=1, \mathbf{b}_{1:w-1}\right)
	\end{align}
	and 
	\begin{align}\label{prm2}
		\mathrm{Pr_m}\!\!\left(\!\hat{b}_w\!=\!1|b_w\!=\!0, \mathbf{b}_{1:w-1}\!\!\right)\!=\!1\!-\!\mathrm{Pr}\!\left(N_w^\mathrm{m}\!<\!\xi|b_w\!=\!0, \mathbf{b}_{1:w-1}\!\right),
	\end{align}
	respectively. The BER of the $w$th bit, denoted by $Q_\mathrm{m}[w|\mathbf{b}_{1:w-1}]$, is given by
	\begin{align}\label{Qm}
		Q_\mathrm{m}[w|\mathbf{b}_{1:w-1}]=&P_1\mathrm{Pr}_\mathrm{m}\left(\hat{b}_w=0|b_w=1, \mathbf{b}_{1:w-1}\right)\notag\\&+P_0\mathrm{Pr}_\mathrm{m}\left(\hat{b}_w=1|b_w=0, \mathbf{b}_{1:w-1}\right).
	\end{align}
	
	We denote $Q_\mathrm{m}$ as the average BER over all realizations of $\mathbf{b}_{1:w-1}$ and all bits from $1$ to $W$. $Q_\mathrm{m}$ is obtained as
	\begin{align}\label{qm}
		Q_\mathrm{m}=\frac{1}{W}\sum_{w=1}^{W}\frac{\sum_{\mathbf{b}_{1:w-1}\in\Psi_w}Q_\mathrm{m}[w|\mathbf{b}_{1:w-1}]}{2^{w-1}}.
	\end{align}
	
	\section{Simulation Framework for MF-based TX}\label{SF}
	In this section, we describe the stochastic simulation framework for the MF-based TX. We use a particle-based simulation method that records the exact position of each vesicle. In particular, we determine coordinates of the release point and release time of each molecule for both scenarios.
	
	For the static scenario, we denote the locations of a vesicle at the start and end of the $\chi$th simulation interval by $(x_{\chi-1}, y_{\chi-1}, z_{\chi-1})$ and $(x_{\chi}, y_\chi, z_\chi)$, respectively. If the distance between a vesicle and the TX's center is larger than $r_{\ss\T}$ at the end of the $\chi$th interval, we assume that the vesicle has hit the TX membrane. As described in Section \ref{MF}, this vesicle then fuses with the TX membrane with probability $k_\mathrm{f}\sqrt{\frac{\pi\Delta t_\mathrm{s}}{D_\mathrm{v}}}$ and is reflected with probability $1-k_\mathrm{f}\sqrt{\frac{\pi\Delta t_\mathrm{s}}{D_\mathrm{v}}}$, where $\Delta t_\mathrm{s}$ is the simulation interval. Molecules stored in a vesicle are released at the time and location where the vesicle is fused with the TX membrane. Thus, we need to derive where and when the fusion of vesicles with the membrane occurred in the simulation. For a vesicle fusing with the TX membrane during the $\chi$th simulation interval, we assume that the intersection between the line that is formed by $(x_{\chi-1}, y_{\chi-1}, z_{\chi-1})$ and $(x_\chi, y_\chi, z_\chi)$ and the membrane is the fusion point whose coordinates, denoted by $(x_{\mathrm{f}, \chi}, y_{\mathrm{f}, \chi}, z_{\mathrm{f}, \chi})$, are given by \cite[eqs. (38)-(40)]{deng2015modeling}
	\begin{align}\label{xr}
		x_{\mathrm{f}, \chi}=\frac{-\Lambda_2\pm\sqrt{\Lambda_2^2-4\Lambda_1\Lambda_3}}{2\Lambda_1},
	\end{align}
	\begin{align}\label{yr}
		y_{\mathrm{f}, \chi}=\frac{(x_{\mathrm{f}, \chi}-x_{\chi-1})(y_\chi-y_{\chi-1})}{x_\chi-x_{\chi-1}}+y_{\chi-1},
	\end{align}
	and
	\begin{align}\label{zr}
		z_{\mathrm{f}, \chi}=\frac{(x_{\mathrm{f}, \chi}-x_{\chi-1})(z_\chi-z_{\chi-1})}{x_\chi-x_{\chi-1}}+z_{\chi-1},
	\end{align}
	respectively, where $\Lambda_1=(x_\chi-x_{\chi-1})^2+(y_\chi-y_{\chi-1})^2+(z_\chi-z_{\chi-1})^2$, $\Lambda_2=2(y_\chi-y_{\chi-1})(x_\chi y_{\chi-1}-x_{\chi-1}y_\chi)+2(z_\chi-z_{\chi-1})(x_\chi z_{\chi-1}-x_{\chi-1}z_\chi)$, and $\Lambda_3=x_{\chi-1}(y_\chi-y_{\chi-1})(x_{\chi-1}y_\chi-2y_{\chi-1}x_\chi+y_{\chi-1}x_\chi)+x_{\chi-1}(z_\chi-z_{\chi-1})(x_{\chi-1}z_\chi-2x_\chi z_{\chi-1}+x_{\chi-1}z_{\chi-1})+(x_\chi-x_{\chi-1})^2(y_{\chi-1}^2+z_{\chi-1}^2-r_{\ss\T})$. In \eqref{xr}, $x_{\mathrm{f}, \chi}$ is chosen by satisfying $(x_{\mathrm{f}, \chi}-x_{\chi-1})(x_{\mathrm{f}, \chi}-x_\chi)<0$ since $x_{\mathrm{f},\chi}$ is between $x_{\chi-1}$ and $x_\chi$. We next determine the release time, which is crucial for the subsequent simulation of the molecule propagation and absorption at the RX, via interpolation.
	We denote $t_{\chi-1}$ as the start time of the $\chi$th simulation interval and $t_{\chi-1}+\Delta t_{\mathrm{f}, \chi}$ as the fusion time of the vesicle within the $\chi$th simulation interval. Since each vesicle follows Brownian motion, the square of the displacement is proportional to the time within the same simulation interval. Therefore, we derive $\Delta t_{\mathrm{f}, \chi}$ as
	\begin{align}\label{tr}
		\Delta t_{\mathrm{f}, \chi}\!\!=\!\!\frac{(x_{\mathrm{f}, \chi}\!-\!x_{\chi-1})^2+(y_{\mathrm{f},\chi}\!-\!y_{\chi-1})^2+(z_{\mathrm{f},\chi}\!-\!z_{\chi-1})^2}{(x_\chi-x_{\chi-1})^2+(y_\chi-y_{\chi-1})^2+(z_\chi-z_{\chi  -1})^2}\Delta t_\mathrm{s}.
	\end{align}
	For vesicles failing to fuse with the TX membrane, we assume in the reflection process that they are sent back to their positions at the start of the current simulation interval \cite{deng2015modeling}.
	
	For the mobile scenario, we denote $(x^\mathrm{m}_{\mathrm{f},\chi}, y^\mathrm{m}_{\mathrm{f},\chi}, z^\mathrm{m}_{\mathrm{f},\chi})$ as coordinates of the fusion point for the mobile TX, which are given by $x_{\mathrm{f},\chi}^\mathrm{m}=x_{\mathrm{f},\chi}+x_{\ss\T}(t_{\chi-1}+\Delta t_{\mathrm{f}, \chi})$, $y_{\mathrm{f},\chi}^\mathrm{m}=y_{\mathrm{f},\chi}+y_{\ss\T}(t_{\chi-1}+\Delta t_{\mathrm{f}, \chi})$, and $z_{\mathrm{f},\chi}^\mathrm{m}=z_{\mathrm{f},\chi}+z_{\ss\T}(t_{\chi-1}+\Delta t_{\mathrm{f}, \chi})$. $x_{\ss\T}(t)$, $y_{\ss\T}(t)$, and $z_{\ss\T}(t)$ are coordinates of the TX's center at time $t$. The calculation of the fusion time for the mobile TX is the same as that for the static TX.
	
	\section{Numerical Results}\label{NR}
	\begin{table*}[!t]
		\centering
		\caption{Simulation Parameters for Numerical Results}\label{tabl}
		{\begin{tabular}{|c|c|c|c|}
				\hline
				\textbf{Parameter}&\textbf{Variable}&\textbf{Value}&\textbf{Reference}\\
				\hline
				Number of vesicles released&$N_\mathrm{v}$&200&\\
				\hline
				Number of molecules stored in each vesicle&$\eta$&5&\\
				\hline
				Radius of the TX/RX&$r_{\ss\T}/r_{\ss\R}\;[\mu\mathrm{m}]$&10&\cite{nakano2019methods}\\
				\hline
				Diffusion coefficient of the vesicle&$D_\mathrm{v}\;[\mu\mathrm{m}^2/\mathrm{s}]$&9, 18, 50&\cite{kyoung2008vesicle}\\
				\hline
				Forward reaction rate&$k_\mathrm{f}\;[\mu\mathrm{m}/\mathrm{s}]$&2, 5, 30&\\
				\hline
				Fixed distance for static MC/Initial distance for mobile MC
				&$l/l_0\;[\mu\mathrm{m}]$&40&\\
				\hline
				Degradation rate of molecule $\sigma$&$k_\mathrm{d}\;[\mathrm{s}^{-1}]$&$0.8$&\cite{deng2015modeling}\\
				\hline
				Diffusion coefficient of molecule $\sigma$&$D_{\sigma}\;[\mu\mathrm{m}^2/\mathrm{s}]$&$1000$&\cite{arjmandi2019diffusive}\\
				\hline
				Time to release vesicles/transmit $b_1$ for mobile MC&$t'\;[\mathrm{s}]$&$2$&
				\\
				\hline
				Diffusion coefficient of the TX/RX for mobile MC&$D_{\ss\T}/D_{\ss\R}\;[\mu\mathrm{m}^2/\mathrm{s}]$&8, 10&\cite{ahmadzadeh2018stochastic}\\
				\hline
				Length of the transmitted binary bits&$W$&5&\\
				\hline
				Bit interval&$\phi\;[\mathrm{s}]$&2&\cite{kuran2011modulation}\\
				\hline
				Probability to transmit bit 1/0&$P_1/P_0$&0.5&\cite{ahmadzadeh2018stochastic}\\
				\hline
				Number of divided release intervals&C&3&\\
				\hline
			\end{tabular}
		}
	\end{table*}
	In this section, we present numerical results to validate our theoretical analysis and provide insightful discussion. The simulation time interval is $\Delta t_\mathrm{s}=0.001\;\mathrm{s}$ and all results are averaged over $10^4$ realizations. Throughout this section, we set simulation parameters as in Table \ref{tabl}, unless otherwise stated. In Figs. 5-8, we vary the forward reaction rate and vesicle diffusion coefficient to investigate their impacts on molecule release, time to reach the peak molecule release probability, and molecule absorption for the static and mobile scenarios. Analytical results in Fig. \ref{term} and Figs. 7-9 are verified with simulations. Specifically, we observe agreement between our simulation results and analytical curves derived in Sections III-V, which demonstrates the accuracy of our analysis.
	\subsection{MF-based TX Model Analysis}
	\begin{figure}[!t]
		\centering	
		\subfigure[Release Probability]{
			\begin{minipage}[t]{1\linewidth}
				\centering
				\includegraphics[width=3in]{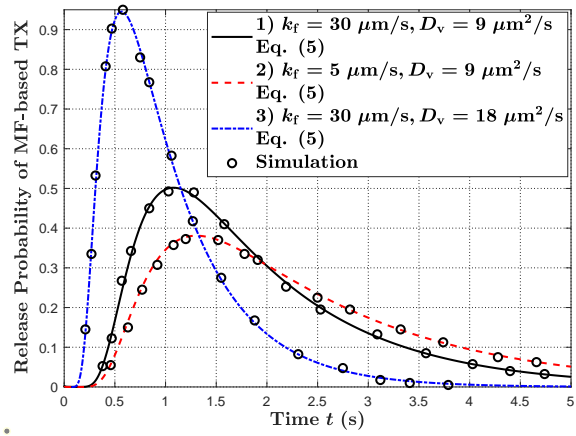}
				\label{a}
			\end{minipage}%
		}%
		\quad
		\subfigure[Number of molecules released]{
			\begin{minipage}[t]{1\linewidth}
				\centering
				\includegraphics[width=3in]{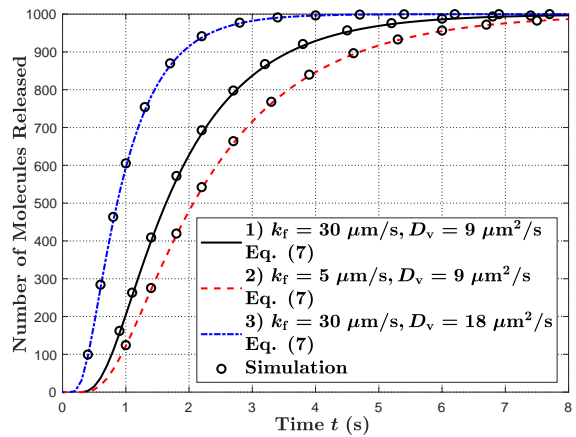}
				\label{b}
			\end{minipage}%
		}
		\centering
		\caption{Molecule release probability from the MF-based TX at time $t$ and the number of molecules released from the MF-based TX by time $t$ versus time $t$ for three parameter sets.}\label{term}
	\end{figure}
	
	In Fig. \ref{term}, we plot the molecule release probability from the TX at time $t$ versus time $t$ in Fig. 5(a) and the number of molecules released from the TX by time $t$ versus time $t$ in Fig. 5(b). First, by comparing parameter sets 1) and 2) in Fig. 5(a), we observe that the peak release probability decreases and the tail of the release probability becomes longer with a decrease in $k_\mathrm{f}$. This is because the decrease in $k_\mathrm{f}$ reduces the fusion probability between a vesicle and the TX membrane. Second, by comparing parameter sets 1) and 3) in Fig. 5(a), we observe that the peak release probability increases and less time is required for the TX to start releasing molecules with an increase in $D_\mathrm{v}$. This is because increasing $D_\mathrm{v}$ accelerates the movement of vesicles. Third, in Fig. 5(b), we observe that all molecules can be released from the TX due to the impulsive release of vesicles if the release period is sufficiently long.
	
	\begin{figure}[!t]
		\begin{center}
			\includegraphics[height=2.2in,width=0.86\columnwidth]{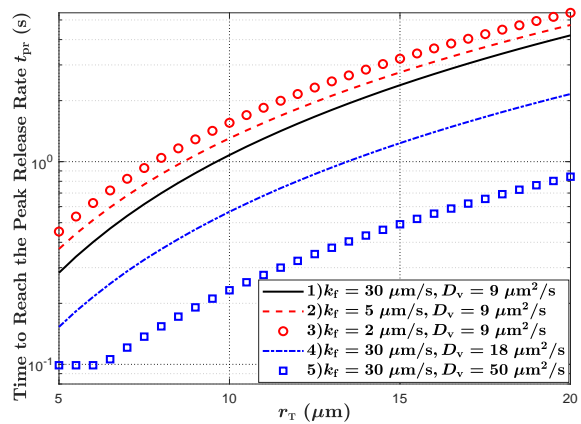}
			\caption{Time to reach the peak molecule release probability from the TX versus $r_{\ss\T}$ for five parameter sets.}\label{hig}\vspace{-0.5em}
		\end{center}
		\vspace{-4mm}
	\end{figure}
	
	In Fig. \ref{hig}, we plot the time to reach the peak molecule release probability from the TX, denoted by $t_\mathrm{pr}$, versus the radius of the TX by searching for the peak value in \eqref{ff} and recording the corresponding time. First, we observe that the time to reach the peak release probability increases with an increase in $r_{\ss\T}$. This is because a vesicle needs to diffuse for a longer time to arrive at a TX membrane with a larger radius. Second, by comparing parameter sets 1), 2), and 3), we observe that the time increases with a decrease in $k_\mathrm{f}$. This is because lower $k_\mathrm{f}$ reduces the fusion probability, such that it takes a longer time to reach the peak probability. Third, by comparing parameter sets 1), 4), and 5), we observe that the time decreases with an increase in $D_\mathrm{v}$. This is because a larger value of $D_\mathrm{v}$ enables vesicles to move faster, such that less time is required for the vesicle to arrive at the TX membrane. 
	
	\subsection{Channel Impulse Response for Static \& Mobile MC}
	\begin{figure}[!t]
		\centering	
		\subfigure[Hitting Probability]{
			\begin{minipage}[t]{1\linewidth}
				\centering
				\includegraphics[width=3in]{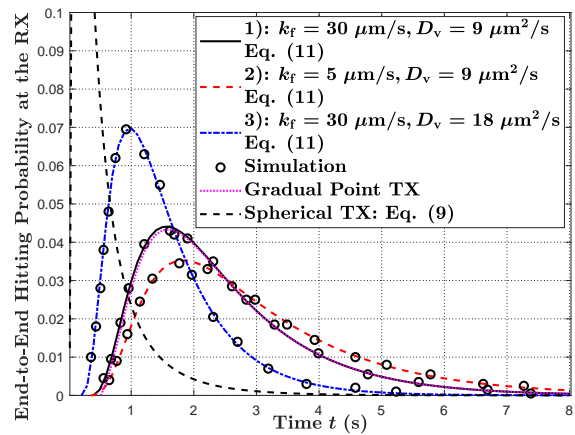}
			\end{minipage}%
		}%
		\quad
		\subfigure[The number of molecules absorbed]{
			\begin{minipage}[t]{1\linewidth}
				\centering
				\includegraphics[width=3in]{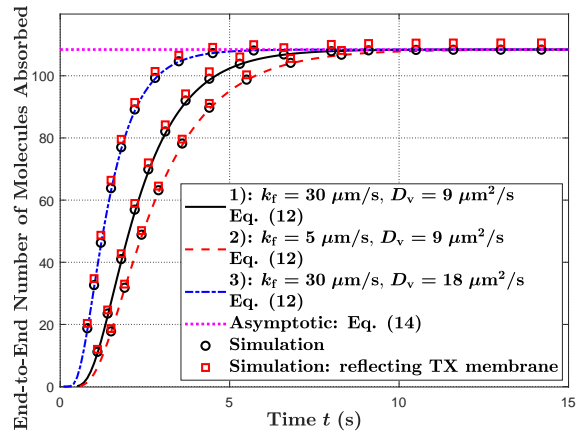}
			\end{minipage}%
		}
		\centering
		\caption{End-to-end molecule hitting probability at the RX at time $t$ and end-to-end number of molecules absorbed at the RX by time $t$ versus time $t$ for three parameter sets.}\label{hr}
	\end{figure}
	\subsubsection{Static MC}
	In Fig. \ref{hr}, we plot the end-to-end molecule hitting probability at the RX at time $t$ versus time $t$ in Fig. 7(a) and the end-to-end number of molecules absorbed at the RX by time $t$ versus time $t$ in Fig. 7(b). We observe several similar trends between the hitting probability in Fig. 7(a) and the release probability in Fig. 5(a) for varying $k_\mathrm{f}$ and $D_\mathrm{v}$. For example, the peak end-to-end hitting probability decreases with a decrease in $k_\mathrm{f}$, as seen when comparing parameter sets 1) and 2), and the peak end-to-end hitting probability increases and less time is required for molecules to start hitting the RX with an increase in $D_\mathrm{v}$, as seen when comparing parameter sets 1) and 3). These similar observations are due to the fact that absorption of molecules at the RX is directly influenced by the release of molecules at the TX. 
	
	The MF-based TX relaxes two major assumptions of the widely-applied point TX model. The first assumption is the instantaneous release of molecules, and the second assumption is molecules released from a point. To investigate the relative importance of these two assumptions to the difference between the MF-based TX and the point TX, we plot the hitting probability at the RX for a gradual point TX and a spherical TX, separately, in Fig 7(a) when $D_\mathrm{v}=9\;\mu\mathrm{m}^2/\mathrm{s}$ and $k_\mathrm{f}=30\;\mu\mathrm{m}/\mathrm{s}$. The gradual point TX is a point TX that releases molecules gradually with the release probability $f_\mathrm{r}(t)$. The hitting probability of the gradual point TX is given by $f_\mathrm{r}(t)*p_\alpha(t)\big|_{l_\alpha=l}$. The spherical TX is a TX that instantaneously releases molecules uniformly over the membrane. The hitting probability of a spherical TX is given in \eqref{pbt}. From Fig. 7(a), we observe the deviation between curves of the MF-based TX and the gradual point TX is extremely small, while the deviation between the MF-based TX and the spherical TX is huge. Therefore, we conclude that the gradual release is the more important element to distinguish between the MF-based TX and the point TX.

	In Fig. 7(b), we observe that the asymptotic number of molecules absorbed at the RX for the three parameter sets is the same. This is because varying $k_\mathrm{f}$ and $D_\mathrm{v}$ do not change the total number of released molecules from the TX after a sufficiently long time as shown in Fig. 5(b), which leads to the total number of molecules absorbed eventually being the same. This observation also complies with Remark \ref{asf}. 
	
	\begin{table*}[!t]
		\centering
		\caption{$R^2$ of $P_\mathrm{e}(t)$ when the TX membrane is reflecting}\label{tab31}
		{\begin{tabular}{|c|c|c|c|c|c|c|c|c|c|c|c|}
				\hline
				$l\;(\mu\mathrm{m})$&40&38&36&34&32&28&26&24&22&21.5&21\\
				\hline
				$R^2$&0.9952&0.9962&0.9964&0.9974&0.9977&0.9995&0.999&0.993&0.9684&0.9457&0.9188\\
				\hline
			\end{tabular}
		}
	\end{table*}
	
	In Fig. 7(b), we also consider the TX membrane as a reflecting boundary to the molecules in the propagation environment to relax assumption A5 in Section \ref{MF}, where molecules are reflected back to their positions at the start of the current simulation interval if they hit the TX membrane. We perform this simulation and plot the number of molecules absorbed. We observe that small differences exist between the simulation and analytical curves for the three parameter sets due to A5. Moreover, we calculate $R^2$. For parameter sets 1), 2) and 3), the $R^2$ is calculated as 0.9952, 0.9980, and 0.9999, respectively, which are quite close to 1. To further investigate the impact of decreasing distance between the TX and RX on $R^2$ for the end-to-end number of molecules absorbed, we calculate $R^2$ for varying $l$ in Table \ref{tab31}, where $k_\mathrm{f}=30\;\mu\mathrm{m}/\mathrm{s}$ and $D_\mathrm{v}=9\;\mu\mathrm{m}^2/\mathrm{s}$. From this Table, we observe that the $R^2$ first increases with the decrease in $l$, reaches the highest at $l=28\;\mu\mathrm{m}$, and then decreases. Compared to a transparent TX membrane, the near side of the reflecting membrane (i.e., closer to the RX) impedes molecules from diffusing further away from the RX such that the RX can absorb more molecules, while the far side of the TX membrane impedes molecules released from that side from being absorbed by the RX. Hence, the highest $R^2$ is achieved when the effects of the near side and far side on molecular absorption are almost equal. Accordingly, the impact of considering a reflecting TX membrane on the CIR analysis is significant when $l$ is extremely small.
	
	\begin{figure}[!t]
		\centering
		
		\subfigure[Varying $k_\mathrm{f}$ and $D_\mathrm{v}$, $t'=2\;\mathrm{s}$]{
			\begin{minipage}[t]{1\linewidth}
				\centering
				\includegraphics[width=3in]{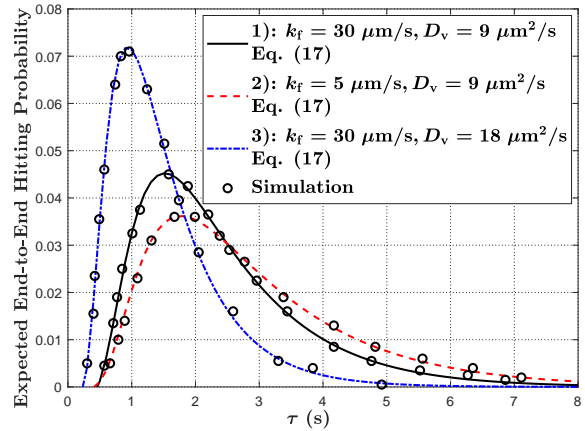}
			\end{minipage}%
		}%
		\quad
		\subfigure[Varying $t'$]{
			\begin{minipage}[t]{1\linewidth}
				\centering
				\includegraphics[width=3in]{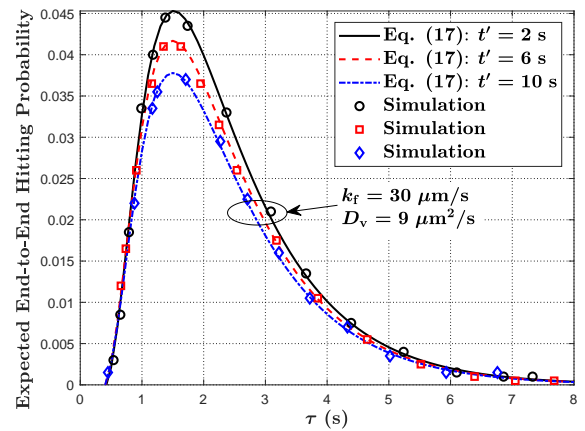}
			\end{minipage}%
		}
		\centering
		\caption{Expected end-to-end molecule hitting probability at the mobile RX at time $t'+\tau$ versus time $\tau$, where $D_{\ss\mathrm{T}}=D_{\ss\mathrm{R}}=8\;\mu\mathrm{m}^2/\mathrm{s}$.}\label{ehr}
	\end{figure}
	\subsubsection{Mobile MC}
	In Fig. \ref{ehr}, we plot the expected end-to-end molecule hitting probability at the mobile RX at time $t'+\tau$ versus time $\tau$, where $k_\mathrm{f}$ and $D_\mathrm{v}$ are varied in Fig. 8(a) and $t'$ is varied in Fig. 8(b). First, the overall trend of the expected end-to-end hitting probability in Fig. 8(a) has identical observations as the hitting probability in Fig. 7(a) for varying $k_\mathrm{f}$ and $D_\mathrm{v}$. Second, in Fig. 8(b), we observe that the expected end-to-end hitting probability decreases with an increase in $t'$. This is because larger $t'$ means that the TX releases an impulse of vesicles after the TX and RX have been diffusing for a longer time, such that the distance between the TX and RX becomes longer on average. This observation also indicates that the communication becomes impractical when $t'$ is sufficiently large since the hitting probability eventually tends to 0. 
	
	\subsection{BER Analysis for Static \& Mobile MC}
	\begin{figure}[!t]
		\centering
		
		\subfigure[Static MC]{
			\begin{minipage}[t]{1\linewidth}
				\centering
				\includegraphics[width=3in]{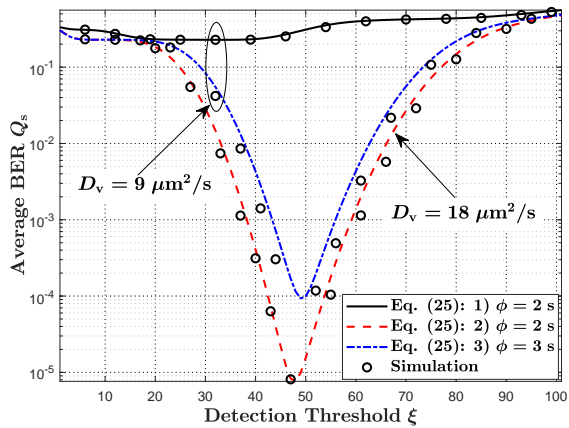}
			\end{minipage}%
		}%
		\quad
		\subfigure[Mobile MC]{
			\begin{minipage}[t]{1\linewidth}
				\centering
				\includegraphics[width=3in]{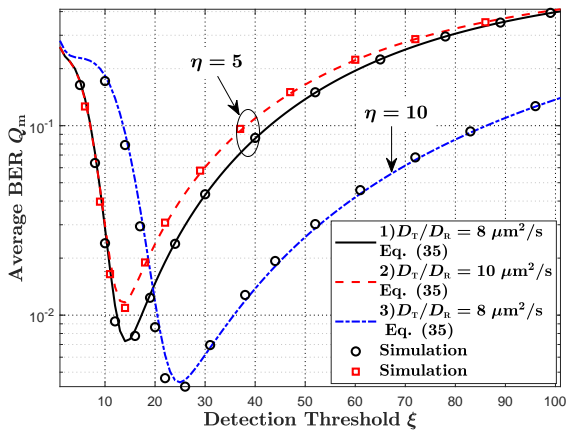}
			\end{minipage}%
		}
		\centering
		\caption{Average BER versus the detection threshold for (a) static MC, where $k_\mathrm{f}=30\;\mu\mathrm{m}/\mathrm{s}$ and (b) mobile MC, where $D_\mathrm{v}=50\;\mu\mathrm{m}^2/\mathrm{s}$ and $k_\mathrm{f}=30\;\mu\mathrm{m}/\mathrm{s}$.}\label{BER}
	\end{figure}
	\subsubsection{Static MC}
	In Fig. 9(a), we plot the average BER $Q_\mathrm{s}$ versus the detection threshold $\xi$ for static MC, where different vesicle diffusion coefficients and bit intervals are adopted to investigate their impacts on the average BER. We first observe that the BER is much larger compared to the other parameter sets when $D_\mathrm{v}=9\;\mu\mathrm{m}^2/\mathrm{s}$ and $\phi=2\;\mathrm{s}$. This is because the period for all molecules to be released from the TX membrane, $\tau_\mathrm{p}$, calculated as $9.5\;\mathrm{s}$, is much larger than the bit interval. Therefore, severe inter-symbol interference (ISI) is expected which dominates the BER. This also illustrates the difference between the MF-based TX and an ideal point TX. For the ideal point TX, the ISI is only caused by the diffusion of molecules in the propagation environment. For the MF-based TX, the ISI is also introduced by the signaling pathways inside the TX. Second, we introduce three methods to decrease the ISI. The first method is increasing the diffusion coefficient of the vesicle. By comparing parameter sets 1) and 2), we observe a dramatic decrease in the BER with the increase in $D_\mathrm{v}$. This is because an increase in $D_\mathrm{v}$ reduces the period for releasing molecules, such that ISI decreases. Given that the active transport of vesicles along microtubules is much faster than free diffusion \cite{ahmed2014active}, the investigation of active transport of vesicles within the TX is an interesting future work for reducing the ISI. The second method is increasing the bit interval. By comparing parameter sets 1) and 3), we observe that the average BER decreases with an increase in $\phi$. This is because a larger $\phi$ enables the RX to absorb more molecules in the current interval, such that fewer molecules are left to influence subsequent bit detection. One drawback of increasing the bit interval is reducing the data transmission rate. The third method is applying a sequence detector in MC to mitigate the ISI as described in \cite{meng2014receiver}.
	\subsubsection{Mobile MC}

	In Fig. 9(b), we plot the average BER $Q_\mathrm{m}$ versus the fixed detection threshold $\xi$ as in \cite{cao2019diffusive} for mobile MC, where different TX and RX diffusion coefficients and the number of molecules stored within each vesicle are applied to investigate their impacts on the average BER. First, by comparing parameter sets 1) and 2), we observe that the average BER increases with an increase in $D_{\ss\T}$ and $D_{\ss\R}$. This is because faster motion of the TX and RX results in a more stochastic channel such that the detection ability at the RX is reduced. Second, by comparing parameter sets 1) and 3), we observe that the minimum BER decreases with an increase in $\eta$. This is because a larger $\eta$ means that more molecules are released from the TX such that more molecules are absorbed within the RX. This increases the detection accuracy at the RX.
	
	\section{Conclusion}\label{con}
	In this paper, we proposed a new TX model that is based on MF between vesicles generated within the TX and the TX membrane to release molecules. We derived the molecule release probability and the fraction of molecules released from the TX. By considering a fully-absorbing RX, we derived the CIR between the MF-based TX and RX when both TX and RX are static or diffusive mobile. By considering a sequence of bits transmitted from the TX, we calculated the average BER for two scenarios. For the mobile scenario, we derived the PMF of the number of molecules absorbed by dividing the molecule release duration into multiple release intervals. Furthermore, we proposed a simulation framework for the MF-based TX. Our numerical results showed that our analytical expressions are accurate. They also showed that a low MF probability or low vesicle mobility slows the release of molecules, extends the time to reach the peak release probability, and reduces the end-to-end molecule hitting probability at the RX. Moreover, numerical results highlighted the difference between the MF-based TX and the ideal point TX in terms of the ISI, since the ISI of the ideal point TX is only caused by the diffusion of molecules in the propagation environment whereas the ISI of the MF-based TX is also caused by the signaling pathways within the TX. Future work includes 1) considering the hindrance, e.g., reflecting or reuptake, of the TX membrane to the diffusion of molecules in the propagation environment and 2) investigating active transport by motor proteins of vesicles along microtubules within the TX.
	\appendices
	\section{Proof of Theorem \ref{theorem1}}\label{app}
	Using separation of variables \cite{cole2010heat}, we express $C(r,t)$ as $C(r,t)=R(r)T(t)$, where $R(r)$ and $T(t)$ are functions of $r$ and $t$, respectively. Substituting $C(r,t)$ into \eqref{fsl}, we obtain
	\begin{align}\label{PDF1}
		\frac{2D_\mathrm{v}R^{'}(r)}{rR(r)}+\frac{D_\mathrm{v}R^{''}(r)}{R(r)}=\frac{T^{'}(t)}{T(t)}\overset{(a)}{=}\hat{\epsilon},
	\end{align}
	where $R^{'}(r)=\frac{\partial R(r)}{\partial r}$, $R^{''}(r)=\frac{\partial^2R(r)}{\partial r^2}$, and $T^{'}(t)=\frac{\partial T(t)}{\partial t}$. The left-hand side of the first equality in \eqref{PDF1} is a function of $r$, denoted by $\hat{\epsilon}(r)$, and the right-hand side is a function of $t$, denoted by $\hat{\epsilon}(t)$. As $\hat{\epsilon}(r)=\hat{\epsilon}(t)$, $\hat{\epsilon}(r)$ and $\hat{\epsilon}(t)$ are equal to a constant $\hat{\epsilon}$ that results in the equality $(a)$. Based on \eqref{PDF1}, we obtain
	\begin{align}\label{Rr}
		r^2R^{''}(r)+2rR^{'}(r)-\frac{\hat{\epsilon}}{D_\mathrm{v}}r^2R(r)=0.
	\end{align}
	
	By setting $\hat{\epsilon}=-D_\mathrm{v}\lambda_n^2$, $n=1,2,3,...$, \eqref{Rr} becomes the radial equation when solving the Helmholtz equation in spherical coordinates \cite{morse1954methods}. For each given $\lambda_n$, the solution of $R(r)$ in \eqref{Rr}, denoted by\footnote{The general solution of the Helmholtz equation is $j_{\hat{n}}(x)$, where $\hat{n}=0,1,2,3,...$. In \eqref{Rr}, $j_0(x)$ is applied since the TX model is symmetric.} $R_n(r)$, is given by $R_n(r)=\mathcal{A}_nj_0(\lambda_nr)$, where $\mathcal{A}_n$ is a constant. As $R_n(r)$ needs to satisfy the boundary condition in \eqref{bc}, we substitute $R_n(r)$ into \eqref{bc} and obtain \eqref{Dvl}. Based on \eqref{PDF1}, we then obtain
	\begin{align}
		\frac{T_n^{'}(t)}{T_n(t)}=-D_\mathrm{v}\lambda_n^2.
	\end{align}
	
	By considering an obvious condition $T_n(t\rightarrow\infty)=0$, one principle solution of $T_n(t)$ is $T_n(t)=B_n\exp\left(-D_\mathrm{v}\lambda_n^2t\right)$. Based on the principle of superposition, $C(r,t)$ becomes
	\begin{align}\label{crt}
		C(r,t)\!=\!\!\sum_{n=1}^{\infty}\!R_n(r)T_n(t)\!=\!\!\sum_{n=1}^{\infty}\!\mathcal{A}_nB_nj_0\!\left(\lambda_nr\right)\!\exp\left(-D_\mathrm{v}\lambda_n^2t\right).
	\end{align}
	
	We next determine $\mathcal{A}_nB_n$ based on the initial condition in \eqref{IC}. According to the Sturm-Liouville theory \cite{pryce1993numerical}, if $n\neq n^{'}$, $rj_0(\lambda_nr)$ and $rj_0(\lambda_{n^{'}}r)$ are orthogonal to each other, which is
	\begin{align}\label{kk}
		\int_{0}^{r_{\ss\T}}r^2j_0(\lambda_nr)j_0(\lambda_{n^{'}}r)\mathrm{d}r=\left\{\begin{array}{lr}
			\frac{2\lambda_nr_{\ss\T}-\mathrm{sin}\left(2\lambda_nr_{\ss\T}\right)}{4\lambda_n^3}, n=n^{'},\\
			0,~~~~~~~~~~~~~~~\;n\neq n^{'}.
		\end{array}
		\right.
	\end{align}
	
	We substitute \eqref{crt} into \eqref{IC} and then multiply $j_0\left(\lambda_nr\right)$ to both sides of the equality. With some mathematical manipulations and taking the integral of both sides with respect to $r$, we obtain
	\begin{align}\label{anbn}
		&\sum_{n=1}^{\infty}\mathcal{A}_nB_n\int_{0}^{r_{\ss\T}}r^2j_0^2(\lambda_nr)\mathrm{d}r=\frac{1}{4\pi}\int_{0}^{r_{\ss\T}}\delta(r)j_0(\lambda_nr)\mathrm{d}r.
	\end{align}
	Based on \eqref{kk} and \eqref{anbn}, we obtain $\mathcal{A}_nB_n=\frac{\lambda_n^3}{\pi(2\lambda_nr_{\ss\T}-\mathrm{sin}(2\lambda_nr_{\ss\T}))}$. By substituting $\mathcal{A}_nB_n$ into \eqref{crt}, we obtain $C(r,t)$. According to \cite[eq. (3.106)]{schulten2000lectures}, the molecule release probability is expressed as $f_\mathrm{r}(t)=4\pi r_{\ss\T}^2k_\mathrm{f}C(r_{\ss\T},t)$. By substituting $C(r_{\ss\T},t)$ into $f_\mathrm{r}(t)$, we obtain \eqref{ff}.
	\section{Proof of Lemma \ref{hrb}}\label{ab}
	We perform the surface integral by considering a small surface element that is the lateral surface of a conical frustum, denoted by $\mathrm{d}S$, on the TX membrane. The point $\alpha$ is on $\mathrm{d}S$ with the coordinates $(x, y, z)$. The base and top radii of this conical frustum are $y$ and $y+\mathrm{d}y$, respectively, and the slant height is $\sqrt{\mathrm{d}^2x+\mathrm{d}^2y}$. Based on the expression for the lateral surface area of a conical frustum, $\mathrm{d}S$ is calculated as $\mathrm{d}S=2\pi y\sqrt{1+\left(\frac{\mathrm{d}y}{\mathrm{d}x}\right)^2}\mathrm{d}x+\pi\frac{\mathrm{d}y}{\mathrm{d}x}\sqrt{1+\left(\frac{\mathrm{d}y}{\mathrm{d}x}\right)^2}\mathrm{d}^2x$. Since $y=\sqrt{r_{\ss\T}^2-x^2}$, we have $\frac{\mathrm{d}y}{\mathrm{d}x}=-\frac{x}{\sqrt{r_{\ss\T}^2-x^2}}$. By substituting $\frac{\mathrm{d}y}{\mathrm{d}x}$ into $\mathrm{d}S$, we obtain $\mathrm{d}S=2\pi r_{\ss\T}\mathrm{d}x-\frac{\pi xr_{\ss\T}}{r_{\ss\T}^2-x^2}\mathrm{d}^2x\approx 2\pi r_{\ss\T}\mathrm{d}x$, where $\frac{\pi xr_{\ss\T}}{r_{\ss\T}^2-x^2}\mathrm{d}^2x$ is omitted since it is a higher order infinitesimal. As $\mathrm{d}S$ is an infinitesimal, we treat the distance between each point on $\mathrm{d}S$ and the center of the RX as $l_\alpha$, where $l_\alpha$ can be expressed based on $x$ as $l_\alpha=\sqrt{r_{\ss\T}^2+l^2-2lx}$. By substituting $l_\alpha$ into \eqref{pa}, we obtain $p_\alpha(t,x)$. Accordingly, the hitting probability at the RX for molecules released from $\mathrm{d}S$ is $\rho p_\alpha(t,x)\mathrm{d}S$. Furthermore, $p_\mathrm{u}(t)$ is obtained by integrating $\rho p_\alpha(t,x)\mathrm{d}S$, i.e., $p_\mathrm{u}(t)=\int_{\Omega_\mathrm{t}}\rho p_\alpha(t,x)\mathrm{d}S=\int_{-r_{\ss\T}}^{r_{\ss\T}}2\pi r_{\ss\T}\rho p_\alpha(t,x)\mathrm{d}x$. By substituting $p_\alpha(t,x)$ into $p_\mathrm{u}(t)$ and solving the integral, we obtain \eqref{pbt}.
	\section{Proof of Corollary \ref{asy}}\label{co}
	According to \eqref{Pv}, we have $P_\mathrm{v}(t)=P_\mathrm{u}(t)*f_\mathrm{r}(t)$. Performing the Laplace transform of this expression, we obtain $\mathcal{P}_\mathrm{v}(s)=\mathcal{F}_\mathrm{r}(s)\mathcal{P}_\mathrm{u}(s)$, where $\mathcal{P}_\mathrm{v}(s)$, $\mathcal{F}_\mathrm{r}(s)$, and $\mathcal{P}_\mathrm{u}(s)$ are the Laplace transform of $P_\mathrm{v}(t)$, $f_\mathrm{r}(t)$, and $P_\mathrm{u}(t)$, respectively. According to the final value theorem \cite[eq. (1)]{chen2007final}, if $P_\mathrm{v}(t)$ has a finite limit as $t\rightarrow\infty$, then we have $\lim\limits_{t\rightarrow\infty}P_\mathrm{v}(t)=\lim\limits_{s\rightarrow 0}s\mathcal{P}_\mathrm{v}(s)$. Thus, we calculate $\lim\limits_{t\rightarrow\infty}P_\mathrm{v}(t)$ as
	\begin{align}\label{fvt}
		&\lim\limits_{t\rightarrow\infty}P_\mathrm{v}(t)=\frac{8r_{\ss\T}^3r_{\ss\R}k_\mathrm{f}\rho\pi}{lD_\mathrm{v}}\sqrt{\frac{D_{\sigma}}{k_\mathrm{d}}}\left[\exp\left(-2\sqrt{\beta_1 k_\mathrm{d}}\right)\right.\notag\\&\left.-\exp\left(-2\sqrt{\beta_2 k_\mathrm{d}}\right)\right]\sum_{n=1}^{\infty}\frac{\lambda_nj_0(\lambda_nr_{\ss\T})}{2\lambda_nr_{\ss\T}-\mathrm{sin}(2\lambda_nr_{\ss\T})}.
	\end{align}
	By substituting $t\rightarrow\infty$ into \eqref{num} and noting that $F_\mathrm{r}(t\rightarrow\infty)=1$, we obtain $	\sum_{n=1}^{\infty}\frac{\lambda_nj_0(\lambda_nr_{\ss\T})}{2\lambda_nr_{\ss\T}-\mathrm{sin}(2\lambda_nr_{\ss\T})}=\frac{D_\mathrm{v}}{4r_{\ss\T}^2k_\mathrm{f}}$. By substituting this expression into \eqref{fvt}, we obtain \eqref{pvasy}.
	
	\bibliographystyle{IEEEtran}
	\bibliography{reff}

\begin{thebibliography}{10}
\providecommand{\url}[1]{#1}
\csname url@samestyle\endcsname
\providecommand{\newblock}{\relax}
\providecommand{\bibinfo}[2]{#2}
\providecommand{\BIBentrySTDinterwordspacing}{\spaceskip=0pt\relax}
\providecommand{\BIBentryALTinterwordstretchfactor}{4}
\providecommand{\BIBentryALTinterwordspacing}{\spaceskip=\fontdimen2\font plus
\BIBentryALTinterwordstretchfactor\fontdimen3\font minus
  \fontdimen4\font\relax}
\providecommand{\BIBforeignlanguage}[2]{{%
\expandafter\ifx\csname l@#1\endcsname\relax
\typeout{** WARNING: IEEEtran.bst: No hyphenation pattern has been}%
\typeout{** loaded for the language `#1'. Using the pattern for}%
\typeout{** the default language instead.}%
\else
\language=\csname l@#1\endcsname
\fi
#2}}
\providecommand{\BIBdecl}{\relax}
\BIBdecl

\bibitem{1907.04239}
X.~Huang, Y.~Fang, A.~Noel, and N.~Yang, ``Membrane fusion-based transmitter
  design for molecular communication systems,'' in \emph{Proc. IEEE ICC 2021},
  Montreal, Canada, Jun. 2021, pp. 1--6.

\bibitem{nakano2013molecular}
T.~Nakano, A.~W. Eckford, and T.~Haraguchi, \emph{Molecular
  {Communication}}.\hskip 1em plus 0.5em minus 0.4em\relax Cambridge University
  Press, 2013.

\bibitem{jamali2019channel}
V.~Jamali, A.~Ahmadzadeh, W.~Wicke, A.~Noel, and R.~Schober, ``Channel modeling
  for diffusive molecular communication--a tutorial review,'' \emph{Proc.
  IEEE}, Jun. 2019.

\bibitem{yilmaz2017chemical}
H.~B. Yilmaz, G.-Y. Suk, and C.-B. Chae, ``Chemical propagation pattern for
  molecular communications,'' \emph{IEEE Wireless Commun. Lett.}, vol.~6,
  no.~2, pp. 226--229, Apr. 2017.

\bibitem{arjmandi2016ion}
H.~Arjmandi, A.~Ahmadzadeh, R.~Schober, and M.~N. Kenari, ``Ion channel based
  bio-synthetic modulator for diffusive molecular communication,'' \emph{IEEE
  Trans. Nanobiosci.}, vol.~15, no.~5, pp. 418--432, Jul. 2016.

\bibitem{schafer2019spherical}
M.~Sch{\"a}fer, W.~Wicke, W.~Haselmayr, R.~Rabenstein, and R.~Schober,
  ``Spherical diffusion model with semi-permeable boundary: A transfer function
  approach,'' in \emph{Proc. IEEE ICC 2020}, Dublin, Ireland, Jun. 2020.

\bibitem{jahn1999membrane}
R.~Jahn and T.~C. S{\"u}dhof, ``Membrane fusion and exocytosis,'' \emph{Annu.
  Rev. Biochem.}, vol.~68, no.~1, pp. 863--911, Jul. 1999.

\bibitem{biomor}
A.~Morgan, ``Exocytosis,'' \emph{Essays BioChem.}, vol.~30, pp. 77--95, 1995.

\bibitem{bonifacino2004mechanisms}
J.~S. Bonifacino and B.~S. Glick, ``The mechanisms of vesicle budding and
  fusion,'' \emph{Cell}, vol. 116, no.~2, pp. 153--166, Jan. 2004.

\bibitem{wu2014exocytosis}
L.-G. Wu, E.~Hamid, W.~Shin, and H.-C. Chiang, ``Exocytosis and endocytosis:
  modes, functions, and coupling mechanisms,'' \emph{Annu. Rev. Physiol.},
  vol.~76, pp. 301--331, Nov. 2013.

\bibitem{nakano2012molecular}
T.~Nakano, M.~J. Moore, F.~Wei, A.~V. Vasilakos, and J.~Shuai, ``Molecular
  communication and networking: Opportunities and challenges,'' \emph{IEEE
  Trans. Nanobiosci.}, vol.~11, no.~2, pp. 135--148, May 2012.

\bibitem{de2019drug}
O.~G. De~Jong, S.~A. Kooijmans, D.~E. Murphy, L.~Jiang, M.~J. Evers, J.~P.
  Sluijter, P.~Vader, and R.~M. Schiffelers, ``Drug delivery with extracellular
  vesicles: from imagination to innovation,'' \emph{Acc. Chem. Res.}, vol.~52,
  no.~7, pp. 1761--1770, Jun. 2019.

\bibitem{akyildiz2011nanonetworks}
I.~F. Akyildiz, J.~M. Jornet, and M.~Pierobon, ``Nanonetworks: {A} new frontier
  in communications,'' \emph{Commun. ACM}, vol.~54, no.~11, pp. 84--89, Nov.
  2011.

\bibitem{huang2020parameter}
X.~Huang, Y.~Fang, A.~Noel, and N.~Yang, ``Parameter estimation in a noisy {1D}
  environment via two absorbing receivers,'' in \emph{Proc. IEEE Globecom
  2020}, Taipei, Taiwan (ROC), Dec. 2020, pp. 1--7.

\bibitem{ahmadzadeh2018stochastic}
A.~Ahmadzadeh, V.~Jamali, and R.~Schober, ``Stochastic channel modeling for
  diffusive mobile molecular communication systems,'' \emph{IEEE Trans.
  Commun.}, vol.~66, no.~12, pp. 6205--6220, Dec. 2018.

\bibitem{cao2019diffusive}
T.~N. Cao, A.~Ahmadzadeh, V.~Jamali, W.~Wicke, P.~L. Yeoh, J.~Evans, and
  R.~Schober, ``Diffusive mobile {MC} with absorbing receivers: Stochastic
  analysis and applications,'' \emph{IEEE Trans. Mol. Biol. Multi-Scale
  Commun.}, vol.~5, no.~2, pp. 84--99, Nov. 2019.

\bibitem{huang2019statistical}
S.~Huang, L.~Lin, H.~Yan, J.~Xu, and F.~Liu, ``Statistical analysis of received
  signal and error performance for mobile molecular communication,'' \emph{IEEE
  Trans. Nanobiosci.}, vol.~18, no.~3, pp. 415--427, Jul. 2019.

\bibitem{chang2005physical}
R.~Chang, \emph{Physical {Chemistry} for the {Biosciences}}.\hskip 1em plus
  0.5em minus 0.4em\relax Sausalito, CA, USA: Univ. Science Books, 2005.

\bibitem{kyoung2008vesicle}
M.~Kyoung and E.~D. Sheets, ``Vesicle diffusion close to a membrane:
  {Intermembrane} interactions measured with fluorescence correlation
  spectroscopy,'' \emph{Biophys. J.}, vol.~95, no.~12, pp. 5789--5797, Dec.
  2008.

\bibitem{sheetz1987movements}
M.~P. Sheetz, R.~Vale, B.~Schnapp, T.~Schroer, and T.~Reese, ``Movements of
  vesicles on microtubules.'' \emph{Ann. N. Y. Acad. Sci.}, vol. 493, pp.
  409--416, Apr. 1987.

\bibitem{milovanovic2017synaptic}
D.~Milovanovic and P.~De~Camilli, ``Synaptic vesicle clusters at synapses: A
  distinct liquid phase?'' \emph{Neuron}, vol.~93, no.~5, pp. 995--1002, Mar.
  2017.

\bibitem{henkel1996synaptic}
A.~W. Henkel, L.~Simpson, R.~A. Ridge, and W.~J. Betz, ``Synaptic vesicle
  movements monitored by fluorescence recovery after photobleaching in nerve
  terminals stained with {FM1-43},'' \emph{J. Neurosci.}, vol.~16, no.~12, pp.
  3960--3967, Jun. 1996.

\bibitem{cosson2002resident}
P.~Cosson, M.~Amherdt, J.~E. Rothman, and L.~Orci, ``A resident {Golgi} protein
  is excluded from peri-{Golgi} vesicles in {NRK} cells,'' \emph{Proc. Natl.
  Acad. Sci.}, vol.~99, no.~20, pp. 12\,831--12\,834, Oct. 2002.

\bibitem{hong2005snares}
W.~Hong, ``{SNAREs} and traffic,'' \emph{Biochim. Biophys. Acta Mol. Cell
  Res.}, vol. 1744, no.~2, pp. 120--144, Apr. 2005.

\bibitem{hua2001three}
Y.~Hua and R.~H. Scheller, ``Three snare complexes cooperate to mediate
  membrane fusion,'' \emph{PNAS}, vol.~98, no.~14, pp. 8065--8070, Jul. 2001.

\bibitem{madelaine2017simplification}
G.~Madelaine, E.~Tonello, C.~Lhoussaine, and J.~Niehren, ``Simplification of
  reaction networks, confluence and elementary modes,'' \emph{Computation},
  vol.~5, no.~1, p.~14, Mar. 2017.

\bibitem{arjmandi2019diffusive}
H.~Arjmandi, M.~Zoofaghari, and A.~Noel, ``Diffusive molecular communication in
  a biological spherical environment with partially absorbing boundary,''
  \emph{IEEE Trans. Commun.}, vol.~67, no.~10, pp. 6858--6867, Jul. 2019.

\bibitem{ahmadzadeh2016comprehensive}
A.~Ahmadzadeh, H.~Arjmandi, A.~Burkovski, and R.~Schober, ``Comprehensive
  reactive receiver modeling for diffusive molecular communication systems:
  Reversible binding, molecule degradation, and finite number of receptors,''
  \emph{IEEE Trans. Nanobiosci.}, vol.~15, no.~7, pp. 713--727, Oct. 2016.

\bibitem{andrews2004stochastic}
S.~S. Andrews and D.~Bray, ``Stochastic simulation of chemical reactions with
  spatial resolution and single molecule detail,'' \emph{Phys. Biol.}, vol.~1,
  no.~3, p. 137, 2004.

\bibitem{andrews2009accurate}
S.~S. Andrews, ``Accurate particle-based simulation of adsorption, desorption
  and partial transmission,'' \emph{Phys. Biol.}, vol.~6, no.~4, p. 046015,
  Nov. 2009.

\bibitem{haluska2006time}
C.~K. Haluska, K.~A. Riske, V.~Marchi-Artzner, J.-M. Lehn, R.~Lipowsky, and
  R.~Dimova, ``Time scales of membrane fusion revealed by direct imaging of
  vesicle fusion with high temporal resolution,'' \emph{PNAS}, vol. 103,
  no.~43, pp. 15\,841--15\,846, Oct. 2006.

\bibitem{jamali2017symbol}
V.~Jamali, A.~Ahmadzadeh, and R.~Schober, ``Symbol synchronization for
  diffusion-based molecular communications,'' \emph{IEEE Trans. Nanobiosci.},
  vol.~16, no.~8, pp. 873--887, Dec. 2017.

\bibitem{kuran2020survey}
M.~{\c{S}}. Kuran, H.~B. Yilmaz, I.~Demirkol, N.~Farsad, and A.~Goldsmith, ``A
  survey on modulation techniques in molecular communication via diffusion,''
  \emph{IEEE Commun. Surv. Tutor.}, vol.~23, no.~1, pp. 7--28, Firstquarter
  2020.

\bibitem{dincc2018impulse}
F.~Din{\c{c}}, B.~C. Akdeniz, A.~E. Pusane, and T.~Tugcu, ``Impulse response of
  the molecular diffusion channel with a spherical absorbing receiver and a
  spherical reflective boundary,'' \emph{IEEE Trans. Mol. Biol. Multi-Scale
  Commun.}, vol.~4, no.~2, pp. 118--122, Jun. 2018.

\bibitem{berg1993random}
H.~C. Berg, \emph{Random {Walks} in {Biology}}.\hskip 1em plus 0.5em minus
  0.4em\relax Princeton, NJ, USA: Princeton Univ. Press, 1993.

\bibitem{schulten2000lectures}
K.~Schulten and I.~Kosztin, \emph{Lectures in {Theoretical}
  {Biophysics}}.\hskip 1em plus 0.5em minus 0.4em\relax Univ. Illinois,
  Champaign, IL, USA, 2000, vol. 117.

\bibitem{olver1960bessel}
F.~W. Olver and L.~C. Maximon, \emph{Bessel {Functions}}.\hskip 1em plus 0.5em
  minus 0.4em\relax New York, NY: Cambridge Univ. Press, 1960.

\bibitem{heren2015effect}
A.~C. Heren, H.~B. Yilmaz, C.-B. Chae, and T.~Tugcu, ``Effect of degradation in
  molecular communication: Impairment or enhancement?'' \emph{IEEE Trans. Mol.
  Biol. Multi-Scale Commun.}, vol.~1, no.~2, pp. 217--229, Jun. 2015.

\bibitem{Miller1958Generalized}
K.~S. Miller, R.~I. Bernstein, and L.~E. Blumenson, ``Generalized {Rayleigh}
  processes,'' \emph{Quart. Appl. Math.}, vol.~16, no.~2, pp. 137--145, Jul.
  1958.

\bibitem{2017Diffusive}
A.~Ahmadzadeh, V.~Jamali, A.~Noel, and R.~Schober, ``Diffusive mobile molecular
  communications over time-variant channels,'' \emph{IEEE Commun. Lett.},
  vol.~21, no.~6, pp. 1265--1268, Jun. 2017.

\bibitem{le1960approximation}
L.~Le~Cam, ``An approximation theorem for the {Poisson} binomial
  distribution.'' \emph{Pac. J. Math}, vol.~10, no.~4, pp. 1181--1197, Nov.
  1960.

\bibitem{lu2016effect}
Y.~Lu, M.~D. Higgins, A.~Noel, M.~S. Leeson, and Y.~Chen, ``The effect of two
  receivers on broadcast molecular communication systems,'' \emph{IEEE Trans.
  Nanobiosci.}, vol.~15, no.~8, pp. 891--900, Dec. 2016.

\bibitem{huang2020initial}
S.~Huang, L.~Lin, W.~Guo, H.~Yan, J.~Xu, and F.~Liu, ``Initial distance
  estimation and signal detection for diffusive mobile molecular
  communication,'' \emph{IEEE Trans. NanoBiosci.}, Jul. 2020.

\bibitem{akdeniz2020equilibrium}
B.~C. Akdeniz, M.~Egan, and B.~Tang, ``Equilibrium signaling: Molecular
  communication robust to geometry uncertainties,'' \emph{IEEE Trans. Commun.},
  pp. 1--1, Oct. 2020.

\bibitem{petrov2012sums}
V.~V. Petrov, \emph{Sums of {Independent} {Random} {Variables}}.\hskip 1em plus
  0.5em minus 0.4em\relax New York, NY, USA: Springer-Verlag, 1975.

\bibitem{deng2015modeling}
Y.~Deng, A.~Noel, M.~Elkashlan, A.~Nallanathan, and K.~C. Cheung, ``Modeling
  and simulation of molecular communication systems with a reversible
  adsorption receiver,'' \emph{IEEE Trans. Mol. Biol. Multi-Scale Commun.},
  vol.~1, no.~4, pp. 347--362, Dec. 2015.

\bibitem{nakano2019methods}
T.~Nakano, Y.~Okaie, S.~Kobayashi, T.~Hara, Y.~Hiraoka, and T.~Haraguchi,
  ``Methods and applications of mobile molecular communication,'' \emph{Proc.
  IEEE}, vol. 107, no.~7, pp. 1442--1456, Jun. 2019.

\bibitem{kuran2011modulation}
M.~S. Kuran, H.~B. Yilmaz, T.~Tugcu, and I.~F. Akyildiz, ``Modulation
  techniques for communication via diffusion in nanonetworks,'' in \emph{Proc.
  IEEE ICC 2011}, Kyoto, Japan, Jun. 2011, pp. 1--5.

\bibitem{ahmed2014active}
W.~W. Ahmed and T.~A. Saif, ``Active transport of vesicles in neurons is
  modulated by mechanical tension,'' \emph{Sci. Rep.}, vol.~4, no.~1, pp. 1--7,
  Mar. 2014.

\bibitem{meng2014receiver}
L.-S. Meng, P.-C. Yeh, K.-C. Chen, and I.~F. Akyildiz, ``On receiver design for
  diffusion-based molecular communication,'' \emph{IEEE Trans. Signal
  Process.}, vol.~62, no.~22, pp. 6032--6044, Nov. 2014.

\bibitem{cole2010heat}
K.~Cole, J.~Beck, A.~Haji-Sheikh, and B.~Litkouhi, \emph{Heat {Conduction}
  {Using} {Greens} {Functions}}.\hskip 1em plus 0.5em minus 0.4em\relax Boca
  Raton, FL, USA: CRC Press, 2010.

\bibitem{morse1954methods}
P.~M. Morse and H.~Feshbach, \emph{Methods of {Theoretical} {Physics}, {Part
  I}}.\hskip 1em plus 0.5em minus 0.4em\relax New York: McGraw-Hill Book Co.,
  1953.

\bibitem{pryce1993numerical}
J.~D. Pryce, \emph{Numerical {Solution} of {Sturm}-{Liouville} {Problems}},
  Oxford, U.K., 1993.

\bibitem{chen2007final}
J.~Chen, K.~H. Lundberg, D.~E. Davison, and D.~S. Bernstein, ``The final value
  theorem revisited-infinite limits and irrational functions,'' \emph{IEEE
  Control Syst. Mag.}, vol.~27, no.~3, pp. 97--99, May 2007.

\end{thebibliography}
\end{document}